\documentclass[%
 reprint,
 amsmath,amssymb,
 aps,
]{revtex4-2}

\usepackage{amsfonts}

\usepackage[caption=false]{subfig} 
\usepackage{graphicx}
\usepackage{dcolumn}
\usepackage{bm}

\usepackage{tikz}
\usepackage[compat=1.1.0]{tikz-feynman}
\usepackage{feynmp}

\newcommand{\dbar}{d\hspace*{-0.08em}\bar{}\hspace*{0.1em}}
\def\deltabar{{\mathchar '26\mkern -10mu\delta}}

\usepackage{bbold}

\usepackage{braket}

\usepackage[hidelinks,colorlinks=true,linkcolor=blue,citecolor=blue]{hyperref}

\DeclareMathOperator{\csch}{csch}

\usepackage[T1]{fontenc}

\usepackage{float}

\begin{document}
\preprint{APS/123-QED}
\hypersetup{urlcolor=black}

\title{Exact solution of a boundary tumbling particle system in one dimension}

\author{Connor Roberts}
 \email{connor.roberts16@imperial.ac.uk}
 
\author{Gunnar Pruessner}
\affiliation{Department of Mathematics, Imperial College London, 180 Queen's Gate, London SW7 2AZ.\\
 Centre for Complexity Science, Imperial College London, UK.
}

\date{\today}

\begin{abstract}
We derive the fully time-dependent solution to a run-and-tumble model for a particle which has tumbling restricted to the boundaries of a one-dimensional interval. This is achieved through a field-theoretic perturbative framework by exploiting an elegant underlying structure of the perturbation theory. We calculate the particle densities, currents and variance as well as characteristics of the boundary tumbling. The analytical findings, in agreement with Monte-Carlo simulations, show how the particle densities are linked to the scale of diffusive fluctuations at the boundaries. The generality of our approach suggests it could be readily applied to similar problems described by Fokker-Planck equations containing localised reaction terms.
\end{abstract}

\maketitle

\section{\label{sec:Intro}Introduction}

Active matter is a class of motile matter whose constituent agents operate far from equilibrium by drawing in energy from their environment to perform directed motion \cite{nardini2017entropy, fodor2016far}. The ability of active matter models to characterise living systems across a wide range of scales has, in recent years, made them the cutting edge in the study of non-equilibrium systems \cite{ramaswamy2010mechanics, gompper20202020}.

A well-studied model of active motility is the run-and-tumble (RnT) process \cite{slowman2017exact, sevilla2019stationary}, which is most notably an idealised model of the motion of \textit{E.\ coli} \cite{renadheer2019path, de2004chemotaxis}. The RnT process is a random walk that consists of ballistic self-propelled `runs' interspersed with exponentially distributed changes in direction, known as `tumbles' \cite{tailleur2008statistical}. In dimensions greater than one, each tumble event draws from a continuous distribution of possible directions, similarly to active Brownian motion \cite{Zhang:2021, solon2015active, cates2013active}. However, in one dimension, RnT dynamics simplify significantly to that of a particle switching between two directions of motion \cite{angelani2015run, dhar2019run, garcia2021run, malakar2018steady, razin2020entropy}. In this regime, confined RnT models can possess non-equilibrium steady states that are analytically accessible \cite{dhar2019run, garcia2021run, malakar2018steady, razin2020entropy, frydel2022run, basu2020exact, frydel2021generalized, angelani2017confined}, making them an attractive route to studying non-equilibrium phenomena analytically.

Here, we present a model of confined RnT motion with diffusion for a particle which has tumbling restricted to the boundaries of a one-dimensional system. Studying such a `boundary tumbling' reaction is instrumental to understanding cellular locomotion, as experiments have shown \textit{E.\ coli} adapt to the presence of confining walls by decreasing their tumble rate when within $20~\mu\mathrm{m}$, approximately one run length, of a boundary \cite{molaei2014failed}. This phenomenon can be studied through RnT models with differing tumble rates in the bulk and at the walls \cite{angelani2017confined}, as is the case here.

More generally, by studying this model, we develop techniques that are generally applicable to solving Fokker-Planck equations containing localised reactions terms \cite{risken1996fokker}. Deriving the solution to this model presents new challenges not encountered in models with spatially uniform tumbling, such as the standard confined RnT model \cite{malakar2018steady, razin2020entropy}. In particular, the incorporation of a localised reaction into a model of diffusive particles turns out to be one of the most significant challenges addressed by the present work. Incorporating a localised boundary reaction is difficult because diffusive fluctuations result in infinitesimally small contact times between the particles and the boundary. Hence, the probability to tumble at the boundaries vanishes if tumbling occurs with a finite rate. This seemingly goes against the everyday coarse-grained perspective where particles can get `stuck' at walls for finite periods of time, such as in bacterial ratchet motors \cite{di2010bacterial}. This coarse-grained perspective is equivalent to a lattice description where the lattice spacing exceeds the typical size of the fluctuations away from the boundary, allowing the particle to reside at an `edge site' for a finite time. This would suggest the reaction probability for a localised reaction does not vanish on a lattice, and thus a particle can still boundary tumble, in contrast with the continuum description. The present work provides a way to prevent the boundary tumbling reaction vanishing in the continuum while still maintaining the correct rate. As we argue below, accounting for the role of diffusion in a boundary reaction is necessary for correctly capturing the physics of finite reaction rates.

Incorporating boundary tumbling simply by imposing boundary conditions is fraught with ambiguity and is difficult to implement mathematically. A more successful method of incorporating diffusion-mediated surface reactions, such as boundary tumbling, is through the concept of boundary local time \cite{grebenkov2020paradigm}, where a reaction occurs when the time the particle spends in the boundary's vicinity exceeds a certain value. Our approach is closely related, yet incorporates the relevant physics directly in the master or Fokker-Planck equation by defining a boundary length scale in a way that prevents the vanishing of the reaction probability in the continuum limit, Appendix \ref{app:LengthScaleDerivation}.

Since the incorporation of boundary reactions in a Fokker-Planck equation draws on a novel framework, we begin in Section~\ref{sec:DriftDiffusion} by solving the simpler Fokker-Planck equation for the reflecting boundaries drift-diffusion system without tumbling. The derivation of the solution to this Fokker-Planck equation sets the stage for subsequent perturbative calculations that introduce the boundary tumbling reaction. The framework we use to perform the perturbative calculations is the path integral in the Doi-Peliti formalism \cite{doi1976second, peliti1985path}. Path integrals allow for the use of field-theoretic techniques, such as Feynman diagrams, that make the perturbative calculations more tractable. The Doi-Peliti formalism is chosen specifically because it respects the particle nature of the system, which is essential to capturing the phenomenology of active systems \cite{bothe2022particle}.

Section~\ref{sec:BoundaryTumbling} discusses the boundary tumbling system and contains the main results of the paper. The most important results are the derivation of closed-form expressions for the fully time-dependent particle density, Eqs.~(\ref{eq:FullPropagatorFourier}) and (\ref{eq:FullPropagatorFourierLeftMover}), and its steady-state limit, Eqs.~(\ref{eq:ProbabilityDensityNeat}) and (\ref{eq:ProbabilityDensityColourless}), which we verify with Monte-Carlo simulations, Fig.~\ref{fig:SimulationComparison}. Obtaining these exact results from a perturbative expansion is possible due to the Feynman diagrams of the field theory having a surprisingly elegant structure that is closed under multiplication. This significant result reduces the calculation of the full perturbation series to a matter of linear algebra, Appendix \ref{app:MathematicalObjects}. The derivation of these solutions demonstrates it is possible to study other physically relevant localised reactions in a confined system, such as bacterial deposition \cite{singh2021guided}, via a perturbative treatment of the corresponding terms in the Fokker-Planck equation.

\section{Drift-diffusion with reflecting boundaries}\label{sec:DriftDiffusion}

\subsection{Model}

We consider a particle moving at position $x \in [0,L]$ on a 1D interval with reflecting boundary conditions at $x=0$ and $x=L$. Reflecting boundary conditions correspond to vanishing particle current at the boundaries and so ensure particle confinement. Confinement is an essential property of the boundary tumbling system and so is implemented at the bare level so that it will be present at every subsequent level of the perturbation theory about small tumble rate, Section~\ref{sec:BoundaryTumbling}.

In the bulk of the interval $x \in (0,L)$ the stochastic motion of the particle is governed by the Langevin equation
\begin{equation}\label{eq:Langevin}
    \dot{x}(t) = v + \sqrt{2D}\eta(t),
\end{equation}
where $\dot{x}$ is the instantaneous velocity of the particle, $t$ is the time, $v$ is the particle's self-propulsion speed, $D$ is the diffusion constant and $\eta(t)$ is a Gaussian white noise defined by $\langle \eta(t) \rangle = 0$ and $\langle \eta(t)\eta(t')\rangle = \delta(t-t')$.

Alternatively, the particle density $P_{0}(x,t)$ evolves deterministically according to the drift-diffusion Fokker-Planck equation \cite{risken1996fokker}
\begin{equation}\label{eq:FokkerPlanckNoTumbling}
    \frac{\partial P_{0}(x,t)}{\partial t} = \mathcal{L}P_{0}(x,t) \equiv \left(D\frac{\partial^{2}}{\partial x^2}- v\frac{\partial}{\partial x}\right)P_{0}(x,t),
\end{equation}
where $\mathcal{L}$ is the drift-diffusion operator. Eq.~(\ref{eq:FokkerPlanckNoTumbling}) can be rewritten in the form of a continuity equation
\begin{equation}\label{eq:ConservationEquation}
    \frac{\partial P_{0}(x,t)}{\partial t} + \frac{\partial J_{0}(x,t)}{\partial x} = 0,
\end{equation}
where the particle current is given by 
\begin{equation}\label{eq:CurrentNoTumbling}
    J_{0}(x,t) = vP_{0}(x,t) - D\frac{\partial P_{0}(x,t)}{\partial x},
\end{equation}
and the reflecting boundary conditions impose $J_{0}(0,t)=J_{0}(L,t)=0$.

\subsection{Solution and field theory}

In order to solve the boundary tumbling system, we first cast the system of equations governing the drift-diffusion system, Eqs.~(\ref{eq:FokkerPlanckNoTumbling}) to (\ref{eq:CurrentNoTumbling}), as a Doi-Peliti field theory \cite{doi1976second, peliti1985path}. This will allow for perturbative calculations in the boundary tumbling, Section~\ref{sec:BoundaryTumbling}, by assigning the Fourier-space representation of the drift-diffusion particle density as the bare propagator in the field theory with boundary tumbling.

We begin by defining the annihilator field
\begin{subequations}\label{eq:FieldDefinitions}
\begin{equation}\label{eq:ProbingField}
    \phi(x,t) = \frac{1}{L}\sum_{n=0}^{\infty}\int \mathrm{\dbar}\omega~ \phi_{n}(\omega)e^{-\mathring{\imath}\omega t}U_{n}(x),
\end{equation}
and the Doi-shifted creation field \cite{cardy2008non}
\begin{equation}\label{eq:CreationField}
    \tilde{\phi}(x,t) = \frac{1}{L}\sum_{m=0}^{\infty}\int \mathrm{\dbar}\omega_{0}~ \tilde{\phi}_{m}(\omega_{0})e^{-\mathring{\imath}\omega_{0} t}\tilde{U}_{m}(x),
\end{equation}
\end{subequations}
where $\mathrm{\dbar} \equiv \mathrm{d}/2\pi$ and $U_n(x)$ and $\tilde{U}_{m}(x)$ are right and left eigenfunctions of $\mathcal{L}$, Eq.~(\ref{eq:FokkerPlanckNoTumbling}), respectively.

By solving Eqs.~(\ref{eq:FokkerPlanckNoTumbling}) to (\ref{eq:CurrentNoTumbling}), Appendix \ref{app:Eigenstates}, the stationary $n=0$ right eigenfunction is found to be the Boltzmann solution
\begin{equation}\label{eq:StationaryRightEigenstateNoTumbling}
    U_{0}(x) = e^{\frac{v}{D}x},
\end{equation}
with eigenvalue $\lambda_{0} = 0$, whilst the $n \in \mathbb{N}^{+}$ eigenfunctions are given by
\begin{equation}\label{eq:RightEigenstatesNoTumbling}
   U_{n}(x) = 2e^{\frac{v}{2D}x}\cos(k_{n}x + \theta_{n}),
\end{equation}
where $k_{n} = n\pi/L$, $\theta_{n} = \arctan(-v/2Dk_{n})$ and the eigenvalues are $\lambda_{n} = -Dk_{n}^{2} - v^{2}/4D$.

The left and right eigenfunctions need to be orthogonal for the operator of the resulting action functional, Eq.~(\ref{eq:ActionNoTumbling}), to be made diagonal and thus rendered local in $n$. This can be achieved by equipping the inner product with a suitable weight function $w(x)$ or, equivalently, by absorbing $w(x)$ into the left eigenfunctions to leave $\mathcal{L}$ in the desired self-adjoint form \cite{horsthemke1984noise}. In the present case, the weight function is $w(x) = \exp(-vx/D)$ \cite{farkas2001one}. Hence, the left eigenfunctions are given by
\begin{equation}\label{eq:StationaryLeftEigenstateNoTumbling}
    \tilde{U}_{0}(x) = \frac{vL/D}{\exp(vL/D)-1},
\end{equation}
\begin{equation}\label{eq:LeftEigenstatesNoTumbling}
   \tilde{U}_{m}(x) = 2e^{-\frac{v}{2D}x}\cos(k_{m}x + \theta_{m}),
\end{equation}
where the specific choice of normalisation
\begin{widetext}
\begin{equation}\label{eq:EigenstatesNormalisation}
   \int_{0}^{L}\mathrm{d}x~\tilde{U}_{m}(x)U_{n}(x) = \begin{cases}
			L, & n=m=0\\
            2L, & n=m>0\\
            0, & n \neq m\\
		 \end{cases}
\end{equation}
results in a simple form for the bare propagator, Eq.~(\ref{eq:BarePropagator}).

By substituting the eigenfunctions, Eqs.~(\ref{eq:StationaryRightEigenstateNoTumbling}) to (\ref{eq:LeftEigenstatesNoTumbling}), into the fields, Eqs.~(\ref{eq:ProbingField}) and (\ref{eq:CreationField}), we determine the action functional for this system to be \cite{garcia2021run, Zhang:2021, bothe2021doi, zhen2022optimal}
\begin{equation}\label{eq:ActionNoTumbling}
\begin{split}
    \mathcal{A}_{0} &= \int\mathrm{d}t\int_{0}^{L}\mathrm{d}x~\tilde{\phi}(x,t)\left(\frac{\partial}{\partial t} - \mathcal{L} \right)\phi(x,t)\\ &=\frac{1}{L}\int\mathrm{\dbar}\omega~\mathrm{\dbar}\omega_{0} ~ \deltabar(\omega+\omega_{0}) \left[\tilde{\phi}_{0}(\omega_{0})~(-\mathring{\imath}\omega + r)~\phi_{0}(\omega) + 2\sum_{n=1}^{\infty}\sum_{m=1}^{\infty}\delta_{nm}~ \tilde{\phi}_{m}(\omega_{0})\left(-\mathring{\imath}\omega + Dk_{n}^{2} + \frac{v^{2}}{4D} + r\right)\phi_{n}(\omega)\right],
    \end{split}
\end{equation}
where $\omega$ is the conjugate Fourier variable to $t$, $\deltabar(\omega+\omega_{0}) \equiv 2\pi \delta(\omega+\omega_{0})$ and $r>0$ is a spontaneous extinction rate which we have introduced to maintain causality and to regularise the infrared divergence at $\omega = 0$ when integrating back to direct time. After performing the path integral, we take $r \rightarrow 0^{+}$ to remove the extinction process.

\begin{figure}
\subfloat[Drift-diffusion particle density $P_{0}(x,t)$, Eq.~(\ref{eq:ProbabilityDensityNoTumbling}).\label{subfig:NoTumblingDensity}]{%
  \includegraphics[width=0.48\columnwidth, trim=1.1cm 0cm 2.6cm 2.4cm, clip]{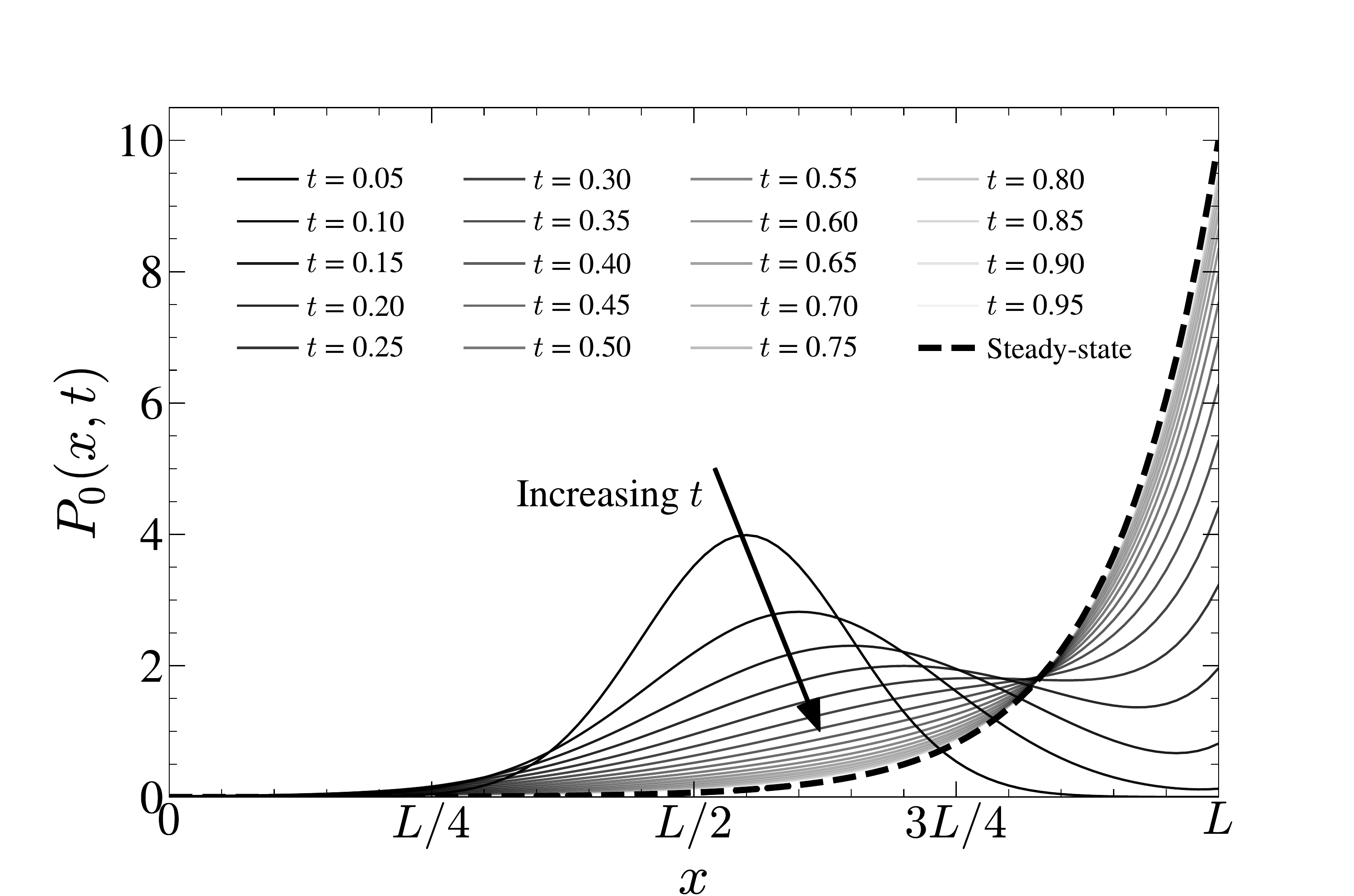}%
}\hfill
\subfloat[Drift-diffusion particle current $J_{0}(x,t)$, Eq.~(\ref{eq:CurrentNoTumbling}).\label{subfig:NoTumblingCurrent}]{%
  \includegraphics[width=0.48\columnwidth, trim=1.1cm 0cm 2.6cm 2.4cm, clip]{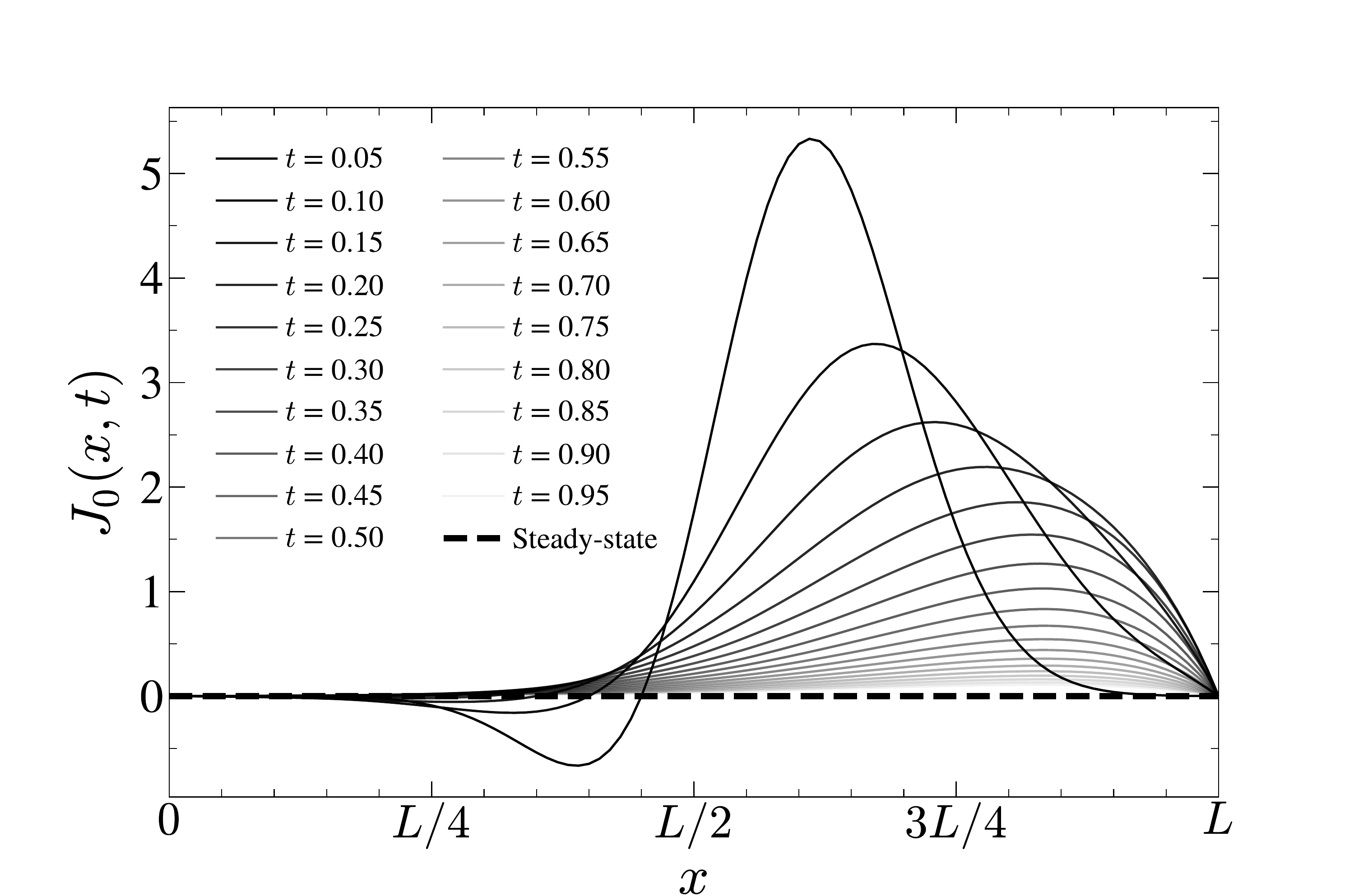}%
}
\caption{Particle density $P_{0}(x,t)$, Eq.~(\ref{eq:ProbabilityDensityNoTumbling}), and current $J_{0}(x,t)$, Eq.~(\ref{eq:CurrentNoTumbling}), for the reflecting boundaries drift-diffusion system on an interval of length $L=1$, initialised at $x_{0}=L/2$ with $v=1$ and $D=0.1$.}\label{fig:DriftDiffusionDensity}
\end{figure}

An observable $\bullet$ is calculated from the bare action via the path integral
\begin{equation}\label{eq:PathIntegralBilinear}
    \langle \bullet \rangle_{0} \equiv \int \mathcal{D}\phi\mathcal{D}\tilde{\phi}~\bullet~e^{-\mathcal{A}_{0}},
\end{equation}
where the subscript $0$ on the expectation value signifies that the integral is performed with respect to the non-perturbative bilinear part of the action $\mathcal{A}_{0}$ that we will perturb about later, see Eq.~(\ref{eq:PathIntegralFull}). The key path integral to perform is for the bare propagator $\langle\phi\tilde{\phi}\rangle_{0}$. The bare propagator can be calculated in closed form after the action has been made local by replacing semi-local derivative operators in $\mathcal{L}$, Eq.~(\ref{eq:FokkerPlanckNoTumbling}), with their algebraic counterparts in Fourier space, Eq.~(\ref{eq:ActionNoTumbling}). Any non-linear terms in the action, and any further linear reaction terms, are incorporated perturbatively, Eq.~(\ref{eq:PathIntegralFull}).

The result of the Gaussian integration in Eq.~(\ref{eq:PathIntegralBilinear}) is that the bare propagator can simply be read off from the action $\mathcal{A}_{0}$, Eq.~(\ref{eq:ActionNoTumbling}), as \cite{le1991quantum}
\begin{equation}\label{eq:BarePropagator}
    \langle \phi_{n}(\omega)\tilde{\phi}_{m}(\omega_{0})\rangle_{0} = \deltabar(\omega+\omega_{0})\frac{\frac{L}{2}\delta_{nm}(1+\delta_{n0})}{-\mathring{\imath}\omega + Dk_{n}^{2} + (1-\delta_{n0})\frac{v^{2}}{4D} + r} \mathrel{\hat{=}} \deltabar(\omega + \omega_{0})\begin{tikzpicture}[dot/.style={fill,circle,inner sep=0pt,outer sep=0pt,minimum size=10pt,label={[label distance=0.3cm]#1}}, baseline={(current bounding box.center)}]
    \begin{feynman}
    \vertex (w) ;
    \vertex [      right=3.5em of w] (w0) ;
    \diagram* {
        (w) -- [red, line width=0.5mm, -] (w0),
    };
    \vertex [above=0.1em of w] {\(n\)};
    \vertex [above=0.1em of w0] {\(m\)};
    \vertex [below=0.5 em of w] {\(\)};
    \end{feynman}
    \end{tikzpicture},
\end{equation}
where $\mathrel{\hat{=}}$ indicates the correspondence between mathematical expressions in the field theory and their Feynman diagrammatic representations, which are read from right to left. Feynman diagrams will be used extensively in Section~\ref{sec:BoundaryTumblingSolution} to make the perturbative calculations more tractable.

The particle density in real space and direct time, i.e.\ the Green's function of Eq.~(\ref{eq:FokkerPlanckNoTumbling}), is calculated from the inverse Fourier transform of the bare propagator,
\begin{equation}\label{eq:InverseFourierTransform}
    P_{0}(x,t|x_{0},t_{0}) \equiv \langle \phi(x,t)\tilde{\phi}(x_{0},t_{0})\rangle_{0} = \frac{1}{L}\sum_{n=0}^{\infty}\frac{1}{L}\sum_{m=0}^{\infty}\int\mathrm{\dbar}\omega~\mathrm{\dbar}\omega_{0}~\langle\phi_{n}(\omega)\tilde{\phi}_{m}(\omega_{0})\rangle_{0}~U_{n}(x)\tilde{U}_{m}(x_{0})~e^{-\mathring{\imath}\omega (t-t_{0})},
\end{equation}
for a particle initialised at $x_{0}$ at time $t_{0}$. By applying the residue theorem, and subsequently taking $r \rightarrow 0^{+}$, the particle density is found to be \footnote{The solution to the reflecting boundaries drift-diffusion equations, Eqs.~(\ref{eq:FokkerPlanckNoTumbling}) to (\ref{eq:CurrentNoTumbling}), first appears in Ref.~\cite{khantha1983diffusion} in Laplace space, and later in various other forms \cite{zhang2009comparison, leroyer2010drift}. The equivalent solution for absorbing boundary conditions is derived in Ref.~\cite{farkas2001one}.}
\begin{equation}\label{eq:ProbabilityDensityNoTumbling}
    P_{0}(x,t|x_{0},t_{0}) = \frac{\Theta(t-t_{0})}{L}\left[\frac{\mathrm{Pe}}{\exp(\mathrm{Pe})-1}e^{\frac{v}{D}x} + 2\sum_{n=1}^{\infty}e^{-\left(Dk_{n}^{2}+\frac{v^{2}}{4D}\right)(t-t_{0})}e^{\frac{v}{2D}(x-x_{0})}\cos(k_{n}x + \theta_{n})\cos(k_{n}x_{0}+\theta_{n}) \right],
\end{equation}
\end{widetext}
where the Heaviside step function $\Theta(t-t_{0})$ incorporates causality, i.e.\ there are no particles in the system until initialisation at time $t = t_{0}$, and the P\'{e}clet number $\mathrm{Pe} = vL/D$ is the ratio of advective to diffusive transport. The drift-diffusion particle density $P_{0}(x,t)$, Eq.~(\ref{eq:ProbabilityDensityNoTumbling}), and its corresponding current $J_{0}(x,t)$, Eq.~(\ref{eq:CurrentNoTumbling}), are evaluated numerically in Fig.~\ref{fig:DriftDiffusionDensity} for a range of times.

\section{Boundary tumbling}\label{sec:BoundaryTumbling}

\subsection{Model}\label{sec:BoundaryTumblingModel}

\begin{figure}
        \centering
        \includegraphics[width=0.48\textwidth, trim = 1.8cm 0.3cm 1.7cm 1.1cm, clip]{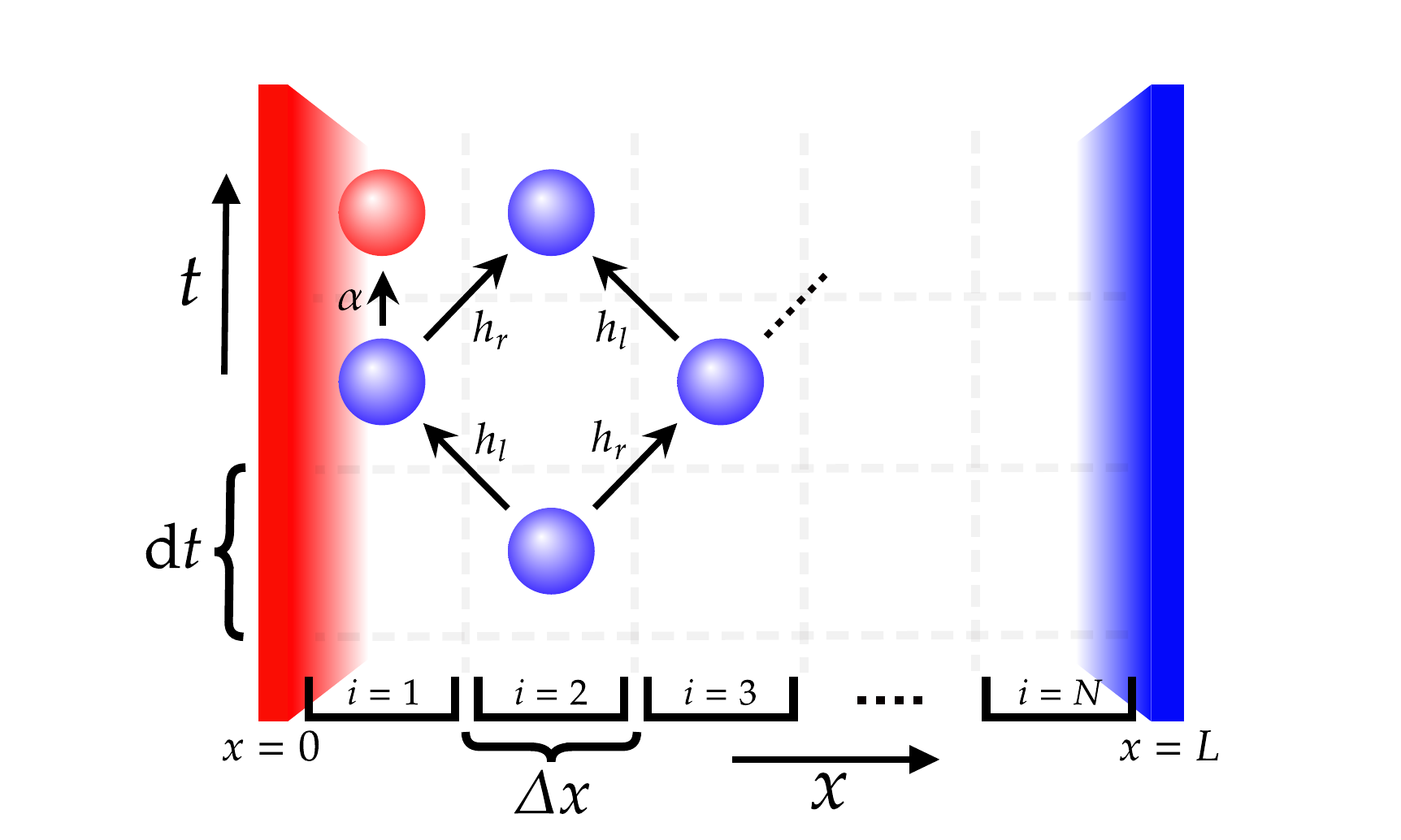}\caption{Schematic representation of the boundary tumbling model on a lattice of $N$ sites. The arrows represent the possible paths a single particle could take. A left-moving particle (blue sphere) hops to the left or to the right in the bulk with respective rates $h_{l} = 2D/\Delta x^{2} + v/\Delta x$ and $h_{r} = 2D/\Delta x^{2} - v/\Delta x$ \cite{van1992stochastic}, where $v>0$ implies a larger left-hopping rate. Upon reaching the left-most lattice site $i=1$, facing the left-hand boundary at $x=0$ (red wall), a left-moving particle converts to a right-moving particle (red sphere) with rate $\alpha$. However, if $\alpha$ remains finite as the continuum limit is taken, the particle's reaction probability vanishes as $\Delta x \rightarrow 0$ because the time the particle spends in the left-most site decreases due to diffusive fluctuations. Thus, as the limit is taken, the reaction rate needs to increase while taking the size $\ell$ of the fluctuations into account. We therefore replace $\alpha$ at the boundary by $\ell\alpha/ \Delta x$, or $\ell\alpha \delta(x)$ in the bulk, Eq.~(\ref{eq:FokkerPlanckCoupledTumbling}). The length scale $\ell$ is given in Eq.~(\ref{eq:DiffusionLengthScale}).}\label{fig:Schematic}
    \end{figure}

The boundary tumbling system is now considered by allowing a particle with positive self-propulsion speed $v>0$ (henceforth referred to as a `right-moving particle') to transmute into a particle with negative self-propulsion speed $-v$, i.e.\ a `left-moving particle', at the right-hand boundary $x=L$, and vice versa at the left-hand boundary $x=0$, whilst retaining all other properties of the system from Section~\ref{sec:DriftDiffusion}. Fig.~\ref{fig:Schematic} shows a schematic representation of the boundary tumbling model.

The boundary tumbling Fokker-Planck equation can be derived rigorously as a continuum limit of the master equation on a lattice. This is the normal route to deriving a Doi-Peliti field theory \cite{cardy2008non, tauber2005applications, bothe2021doi}, specifically by applying the canonical quantisation procedure to rewrite, in terms of ladder operators, the evolution of the probability for a particular occupation number configuration. The continuum limit ultimately recovers the Fokker-Planck equation \footnote{R.\ Garcia-Millian and G.\ Pruessner, To be
published (2022)}, meaning the Doi-Peliti action functional is readily obtained from the Fokker-Planck operator, as in Eq.~(\ref{eq:ActionNoTumbling}). The only additional terms to the drift-diffusion Fokker-Planck equation, Eq.~(\ref{eq:FokkerPlanckNoTumbling}), are the boundary tumbling for both states. As in a master equation, these additional terms are just the normal transmutation terms \cite{garcia2021run, Zhang:2021, zhen2022optimal} supplemented with Kronecker-delta functions to restrict the reaction to the end lattice sites, Appendix \ref{app:LengthScaleDerivation}. In taking the continuum limit, the Kronecker-delta functions become Dirac-delta functions with an accompanying length scale $\ell$ to restore dimensional consistency, discussed in detail below. Hence, the boundary tumbling Fokker-Planck equation is
\begin{subequations}\label{eq:FokkerPlanckCoupledTumbling}
\begin{equation}\label{eq:FokkerPlanckCoupledTumblingRightMovers}
    \frac{\partial P_{R}}{\partial t} =  D\frac{\partial^{2}P_{R}}{\partial x^2}- v\frac{\partial P_{R}}{\partial x} + \ell\alpha\Big[\delta(x)P_{L} - \delta(x-L)P_{R} \Big],
\end{equation}
\begin{equation}\label{eq:FokkerPlanckCoupledTumblingLeftMovers}
    \frac{\partial P_{L}}{\partial t} =  D\frac{\partial^{2}P_{L}}{\partial x^2}+v\frac{\partial P_{L}}{\partial x} - \ell\alpha\Big[\delta(x)P_{L} - \delta(x-L)P_{R} \Big],
\end{equation}
\end{subequations}
where $P_{R/L} = P_{R/L}(x,t)$ is the particle density for right/left-moving particles at position $x$ at time $t$ and $\alpha$ is the tumble rate at which a particle changes its self-propulsion direction at the boundaries.

As explained in Fig.~\ref{fig:Schematic}, the delta functions in Eq.~(\ref{eq:FokkerPlanckCoupledTumbling}) enforce conversion between the different particle states to occur only at the boundaries of the system \footnote{We use the convention $\int_{0}^{L}\mathrm{d}x~\delta(x) = \int_{0}^{L}\mathrm{d}x~\delta(x-L) = 1$.}.\ For instance, the term proportional to $\delta(x)$ corresponds to conversion of a left-moving particle to a right-moving particle at the left-hand boundary at $x=0$. The opposing signs of this term in the coupled equations represent gain and loss to each state respectively. Despite the addition of these singular terms, the solutions to Eq.~(\ref{eq:FokkerPlanckCoupledTumbling}), which are derived in Section~\ref{sec:BoundaryTumblingSolution}, are non-singular and agree with physical expectations and Monte-Carlo simulations.

The standard RnT model with uniform tumble rate is recovered by replacing both $\ell\alpha\delta(x)$ and $\ell\alpha\delta(x-L)$ in Eq.~(\ref{eq:FokkerPlanckCoupledTumbling}) with $\alpha/2$ \cite{razin2020entropy, malakar2018steady}. Thus, by comparing the present boundary RnT model with the standard RnT model, the need for the length scale $\ell$ is apparent on the grounds of dimensionality. However, the physical interpretation of $\ell$ arises from the infinitesimally small contact time between a particle and a boundary due to diffusive fluctuations, Fig.~\ref{fig:Schematic}. Exceptions to this vanishing contact time occur in the absence of diffusive fluctuations $D=0$, or if the dynamics take place on a lattice with lattice spacing $\Delta x$ greater than the typical size $\ell$ of the fluctuations away from the boundary. However, if the dynamics are scrutinised at finer spatial resolution $\Delta x < \ell$, then it becomes apparent that the particle makes an infinite number of microscopically small excursions away from the boundary. As $D$ is increased, more coarse graining, or, equivalently, larger $\Delta x$, is required for a particle to appear to reside at a boundary. Hence, to prevent the boundary tumbling probability from vanishing, $\ell$ in Eq.~(\ref{eq:FokkerPlanckCoupledTumbling}) must increase with $D$ to compensate for the increase in size of the fluctuations. More specifically, we define this length scale such that a particle that resides at the boundary site with a probability according to the steady-state density of a drift-diffusive particle, Section~\ref{sec:DriftDiffusion}, transmutes with average rate $\alpha$. Leaving the details to Appendix \ref{app:LengthScaleDerivation}, where it is derived from the lattice description, the correct scale in the continuum is
\begin{equation}\label{eq:DiffusionLengthScale}
    \ell = \frac{D}{v}\left(1-e^{-\frac{vL}{D}}\right),
\end{equation}
which is approximated by $\ell \approx D/v$ for $D/v \ll L$, and becomes limited by the system size, i.e.\ $\ell \rightarrow L$, for $D/v \gg L$.

\subsection{Solution}\label{sec:BoundaryTumblingSolution}

The Fokker-Planck equation, Eq.~(\ref{eq:FokkerPlanckCoupledTumbling}), couples two partial differential equations by terms proportional to delta functions. In the following, we use a perturbation theory to treat the boundary tumbling terms as perturbations to the bare propagator, Eq.~(\ref{eq:BarePropagator}). Both the left-moving and right-moving states have the same bare propagator since Eq.~(\ref{eq:BarePropagator}) is invariant under $v \rightarrow -v$. However, the eigensystem of the left-moving species differs from that of the right-moving species by the sign of $v$, as can be gleaned from Eqs.~(\ref{eq:FokkerPlanckNoTumbling}) and (\ref{eq:FokkerPlanckCoupledTumblingRightMovers}) compared to (\ref{eq:FokkerPlanckCoupledTumblingLeftMovers}). To this end, we may introduce left-moving eigenfunctions $V_{n}(x)$ and $\tilde{V}_{n}(x)$, defined as in Eqs.~(\ref{eq:StationaryRightEigenstateNoTumbling}) to (\ref{eq:LeftEigenstatesNoTumbling}) by replacing $v$ with $-v$. As a result, $U_{n}$ and $\tilde{V}_{n}$, and similarly $\tilde{U}_{n}(x)$ and $V_{n}(x)$, are generally not orthogonal for $v \neq 0$. Ordinarily, a reaction term would therefore pose a great hurdle, so the localisation of the transmutation due to the delta functions is actually a blessing in disguise.

The perturbative treatment of the boundary tumbling begins with adding the following terms to the action of Eq.~(\ref{eq:ActionNoTumbling}),
\begin{widetext}
\begin{subequations}\label{eq:PerturbativeActions}
\begin{equation}\label{eq:ActionRightToLeft}
    -\mathcal{A}_{R\rightarrow L} = \ell\alpha\int\mathrm{d}t\int_{0}^{L}\mathrm{d}x~\delta(x-L)\left(\tilde{\psi} -  \tilde{\phi}\right)\phi = \int \mathrm{\dbar}\omega_{a} ~ \mathrm{\dbar}\omega_{b} ~\deltabar(\omega_{a} + \omega_{b})\sum_{ij}\Big(\tilde{\psi}_{i}(\omega_{a})\Delta_{\tilde{\psi}_{i}\phi_{j}}+\tilde{\phi}_{i}(\omega_{a})\Delta_{\tilde{\phi}_{i}\phi_{j}} \Big)\phi_{j}(\omega_{b}),
\end{equation}
\begin{equation}\label{eq:ActionLeftToRight}
    -\mathcal{A}_{L\rightarrow R} = \ell\alpha\int\mathrm{d}t\int_{0}^{L}\mathrm{d}x~\delta(x)\left(\tilde{\phi} -  \tilde{\psi}\right)\psi = \int \mathrm{\dbar}\omega_{a} ~ \mathrm{\dbar}\omega_{b} ~\deltabar(\omega_{a} + \omega_{b})\sum_{ij}\Big(\tilde{\phi}_{i}(\omega_{a})\Delta_{\tilde{\phi}_{i}\psi_{j}}+\tilde{\psi}_{i}(\omega_{a})\Delta_{\tilde{\psi}_{i}\psi_{j}} \Big)\psi_{j}(\omega_{b}),
\end{equation}
\end{subequations}
\end{widetext}
where $\psi = \psi(x,t)$ and $\tilde{\psi} = \tilde{\psi}(x,t)$ are, respectively, the annihilation and creation fields of the left-moving particle. The fields of the right-moving particle $\phi$, $\tilde{\phi}$ retain the same definitions from Eq.~(\ref{eq:FieldDefinitions}). Eq.~(\ref{eq:ActionRightToLeft}) is the action for the conversion that occurs at the right-hand boundary, qualitatively representing the creation of a left-moving particle $\tilde{\psi}\phi$ and the annihilation of a right-moving particle $-\tilde{\phi}\phi$. The integrals over eigenfunctions are matrices $\Delta$ in Fourier space. For instance, the matrices
\begin{subequations}\label{eq:TransmutationIntegrals}
\begin{equation}\label{eq:IntegralLeftToRight}
\Delta_{\tilde{\psi}_{i}\phi_{j}} = \frac{\ell\alpha}{L^{2}} \int_{0}^{L}\mathrm{d}x~\delta(x-L)~\tilde{V}_{i}(x)U_{j}(x),
\end{equation}
\begin{equation}\label{eq:IntegralRightToRight}
\Delta_{\tilde{\phi}_{i}\phi_{j}} = -\frac{\ell\alpha}{L^{2}} \int_{0}^{L}\mathrm{d}x~\delta(x-L)~\tilde{U}_{i}(x)U_{j}(x),
\end{equation}
\end{subequations}
arise from the $\tilde{\psi}\phi$ and $\tilde{\phi}\phi$ terms in Eq.~(\ref{eq:ActionRightToLeft}) respectively. The first (respectively, second) index of the matrices $\Delta$ denotes their row (respectively, column) number. The matrix elements are calculated explicitly in Appendix \ref{app:MatrixIntegrals}.

The full action is given by $\mathcal{A} = \mathcal{A}_{0} + \mathcal{A}_{R\rightarrow L} + \mathcal{A}_{L\rightarrow R}$, where the bilinear part of the action $\mathcal{A}_{0}$ now has an extra contribution for the $\psi$ fields, i.e.
\begin{widetext}
\begin{equation}\label{eq:BilinearActionTumbling}
\begin{split}
    \mathcal{A}_{0} = \frac{1}{L}\int\mathrm{\dbar}\omega~\mathrm{\dbar}\omega_{0} ~ \deltabar(\omega+\omega_{0})\Bigg[&\tilde{\phi}_{0}(\omega_{0})~(-\mathring{\imath}\omega + r)~\phi_{0}(\omega) + 2\sum_{n=1}^{\infty}\sum_{m=1}^{\infty}\delta_{nm}~ \tilde{\phi}_{m}(\omega_{0})\left(-\mathring{\imath}\omega + Dk_{n}^{2} + \frac{v^{2}}{4D} + r\right)\phi_{n}(\omega)\\
    &+\tilde{\psi}_{0}(\omega_{0})~(-\mathring{\imath}\omega + r)~\psi_{0}(\omega) + 2\sum_{n=1}^{\infty}\sum_{m=1}^{\infty}\delta_{nm}~ \tilde{\psi}_{m}(\omega_{0})\left(-\mathring{\imath}\omega + Dk_{n}^{2} + \frac{v^{2}}{4D} + r\right)\psi_{n}(\omega)\Bigg].
    \end{split}
\end{equation}
\end{widetext}
Perturbative corrections for $\mathcal{A}_{R\rightarrow L}$ and $\mathcal{A}_{L\rightarrow R}$ are performed by writing the path integral for $\mathcal{A}$ as
\begin{equation}\label{eq:PathIntegralFull}
\begin{split}
    \langle \bullet \rangle &\equiv \int \mathcal{D}\phi\mathcal{D}\tilde{\phi}\mathcal{D}\psi\mathcal{D}\tilde{\psi}~\bullet~e^{-\mathcal{A}_{0}-\mathcal{A}_{R\rightarrow L}-\mathcal{A}_{L\rightarrow R}}\\
    &= \langle \bullet~e^{-\mathcal{A}_{R\rightarrow L}-\mathcal{A}_{L\rightarrow R}} \rangle_{0}\\
    &= \sum_{N=0}^{\infty}\sum_{M=0}^{\infty}\frac{1}{N!}\frac{1}{M!}\langle \bullet~(-\mathcal{A}_{R\rightarrow L})^{N}(-\mathcal{A}_{L\rightarrow R})^{M} \rangle_{0},
    \end{split}
\end{equation}
where the definition of the bilinear path integral, Eq.~(\ref{eq:PathIntegralBilinear}), has been used in the second line and, in the third line, the exponential terms have been expanded as a power series of corrections. Wick's theorem \cite{le1991quantum} can then be used to reduce the products of $\phi$, $\tilde{\phi}$, $\psi$ and $\tilde{\psi}$ on the right-hand side of the final line in Eq.~(\ref{eq:PathIntegralFull}) to products of $\langle\phi\tilde{\phi}\rangle_{0}$ and $\langle\psi\tilde{\psi}\rangle_{0}$, given by Eq.~(\ref{eq:BarePropagator}).

We now proceed to find a closed-form representation of the infinite perturbation series, Eq.~(\ref{eq:PathIntegralFull}), so the result can be converted back to real space. Each term in the series corresponds to a unique Feynman diagram. For instance, the first-order corrections to the particle density for right-moving particles are represented by
\begin{subequations}
\begin{equation}\label{eq:FirstOrderCorrectionsRightRight}
    \langle \phi_{n}(\omega)\tilde{\phi}_{m}(\omega_{0})(-\mathcal{A}_{R\rightarrow L})\rangle_{0} \mathrel{\hat{=}} \deltabar(\omega + \omega_{0})\begin{tikzpicture}[circ/.style={shape=circle, inner sep=2pt, outer sep =0.5pt, line width=0.4mm, draw, node contents=}, baseline={(current bounding box.center)}]
    \begin{feynman}
    \node (m1) ;
    \vertex [left=2 em of m1] (w) ;
    \vertex [right=2 em of m1] (w0) ;
    \diagram* {
        (w) -- [red, line width=0.5mm, -] (m1) -- [red, line width=0.5mm, -] (w0),
    };
    \vertex [circ, below=0em of m1];
    \vertex [above=0.1em of w] {\(n\)};
    \vertex [above=0.1em of w0] {\(m\)};
    \vertex [below=1.2 em of m1] {\(\)};
    \end{feynman}
    \end{tikzpicture},
    \end{equation}
    \begin{equation}\label{eq:FirstOrderCorrectionsRightLeft}
    \langle \phi_{n}(\omega)\tilde{\psi}_{m}(\omega_{0})(-\mathcal{A}_{L\rightarrow R})\rangle_{0} \mathrel{\hat{=}} \deltabar(\omega + \omega_{0})\begin{tikzpicture}[circ/.style={shape=circle, inner sep=2pt, outer sep =0.5pt, line width=0.4mm, draw, node contents=}, baseline={(current bounding box.center)}]
    \begin{feynman}
    \node (m1) ;
    \vertex [left=2 em of m1] (w) ;
    \vertex [right=2 em of m1] (w0) ;
    \diagram* {
        (w) -- [red, line width=0.5mm, -] (m1) -- [blue, line width=0.5mm, -] (w0),
    };
    \vertex [circ, below=0em of m1];
    \vertex [above=0.1em of w] {\(n\)};
    \vertex [above=0.1em of w0] {\(m\)};
    \vertex [below=1.2 em of m1] {\(\)};
    \end{feynman}
    \end{tikzpicture},
\end{equation}
\end{subequations}
depending on whether a right-moving or left-moving particle is initialised respectively. More specifically, Eq.~(\ref{eq:FirstOrderCorrectionsRightLeft}) is a contribution to the particle density of observing a right-moving particle after an initially left-moving particle makes contact with the left-hand boundary exactly once.

The Fourier-space solutions to the boundary tumbling Fokker-Planck equation, Eq.~(\ref{eq:FokkerPlanckCoupledTumbling}), will be referred to as full propagators. Dyson's equation relates a matrix of full propagators to a geometric series over the lower-order corrections in Eq.~(\ref{eq:PathIntegralFull}) \cite{tauber2014critical},
\begin{equation}\label{eq:FeynmanMatrixExpansion}
\begin{split}
\mathsf{F} &\equiv
\begin{pmatrix}
\langle \phi(\omega) \tilde{\phi}(\omega_{0}) \rangle &  \langle \phi(\omega) \tilde{\psi}(\omega_{0}) \rangle \\
 \langle \psi(\omega) \tilde{\phi}(\omega_{0}) \rangle &  \langle \psi(\omega) \tilde{\psi}(\omega_{0}) \rangle
\end{pmatrix}\\
&\mathrel{\hat{=}}
\deltabar(\omega + \omega_{0})\begin{pmatrix}\\[-18pt]
\begin{tikzpicture}[dot/.style={fill,circle,inner sep=0pt,outer sep=0pt,minimum size=8pt,label={[label distance=0cm]#1}}, baseline={(current bounding box.center)}]
    \begin{feynman}
    \node [dot] (m1) ;
    \vertex [   left=2 em of m1] (w) ;
    \vertex [      right=2 em of m1] (w0) ;
    \diagram* {
        (w) -- [red, line width=0.5mm, -] (m1) -- [red, line width=0.5mm, -] (w0),
    };
    \end{feynman}
    \end{tikzpicture} & \begin{tikzpicture}[dot/.style={fill,circle,inner sep=0pt,outer sep=0pt,minimum size=8pt,label={[label distance=0cm]#1}}, baseline={(current bounding box.center)}]
    \begin{feynman}
    \node [dot] (m1) ;
    \vertex [    left=2 em of m1] (w) ;
    \vertex [      right=2 em of m1] (w0) ;
    \diagram* {
        (w) -- [red, line width=0.5mm, -] (m1) -- [blue, line width=0.5mm, -] (w0),
    };
    \end{feynman}
    \end{tikzpicture}\\
\begin{tikzpicture}[dot/.style={fill,circle,inner sep=0pt,outer sep=0pt,minimum size=8pt,label={[label distance=0cm]#1}}, baseline={(current bounding box.center)}]
    \begin{feynman}
    \node [dot] (m1) ;
    \vertex [    left=2 em of m1] (w) ;
    \vertex [      right=2 em of m1] (w0) ;
    \diagram* {
        (w) -- [blue, line width=0.5mm, -] (m1) -- [red, line width=0.5mm, -] (w0),
    };
    \end{feynman}
    \end{tikzpicture} & \begin{tikzpicture}[dot/.style={fill,circle,inner sep=0pt,outer sep=0pt,minimum size=8pt,label={[label distance=0cm]#1}}, baseline={(current bounding box.center)}]
    \begin{feynman}
    \node [dot] (m1) ;
    \vertex [    left=2 em of m1] (w) ;
    \vertex [      right=2 em of m1] (w0) ;
    \diagram* {
        (w) -- [blue, line width=0.5mm, -] (m1) -- [blue, line width=0.5mm, -] (w0),
    };
    \end{feynman}
    \end{tikzpicture}\\[5pt]
\end{pmatrix}\\
&= \deltabar(\omega + \omega_{0})\left(\mathsf{G} + \mathsf{G}\mathsf{T}\mathsf{G} + \mathsf{G}\mathsf{T}\mathsf{G}\mathsf{T}\mathsf{G} + \dots\right)\\
&= \deltabar(\omega + \omega_{0})~\mathsf{G}\left(\mathbb{1} + \mathsf{T}\mathsf{G} + \mathsf{T}\mathsf{G}\mathsf{T}\mathsf{G} + \dots\right)\\
&= \deltabar(\omega + \omega_{0})~\mathsf{G}\left(\mathbb{1} - \mathsf{T}\mathsf{G}\right)^{-1},
    \end{split}
\end{equation}
where
\begin{equation}\label{eq:BarePropagatorMatrix}
    \mathsf{G} \mathrel{\hat{=}}
\begin{pmatrix}
\begin{tikzpicture}[dot/.style={fill,circle,inner sep=0pt,outer sep=0pt,minimum size=10pt,label={[label distance=0.3cm]#1}}, baseline={(current bounding box.center)}]
    \begin{feynman}
    \vertex (w) ;
    \vertex [      right=3em of w] (w0) ;
    \diagram* {
        (w) -- [red, line width=0.5mm, -] (w0),
    };
    \end{feynman}
    \end{tikzpicture} &  0 \\[5pt]
 0 & \begin{tikzpicture}[dot/.style={fill,circle,inner sep=0pt,outer sep=0pt,minimum size=10pt,label={[label distance=0.3cm]#1}}, baseline={(current bounding box.center)}]
    \begin{feynman}
    \vertex (w) ;
    \vertex [      right=3em of w] (w0) ;
    \diagram* {
        (w) -- [blue, line width=0.5mm, -] (w0),
    };
    \end{feynman}
    \end{tikzpicture}
\end{pmatrix}
\end{equation}
is the matrix of bare propagators, Eq.~(\ref{eq:BarePropagator}), and
\begin{equation}\label{eq:TransmutationMatrix}
    \mathsf{T} \equiv 
    \begin{pmatrix}
\Delta_{\tilde{\phi}\phi} &  \Delta_{\tilde{\phi}\psi} \\
\Delta_{\tilde{\psi}\phi} &  \Delta_{\tilde{\psi}\psi}
\end{pmatrix} \mathrel{\hat{=}}
\begin{pmatrix}
\begin{tikzpicture}[circ/.style={shape=circle, inner sep=2pt, outer sep =0.5pt, line width=0.4mm, draw, node contents=}, baseline={(current bounding box.center)}]
    \begin{feynman}
    \node (m1) ;
    \vertex [left=0.8 em of m1] (w) ;
    \vertex [right=0.8 em of m1] (w0) ;
    \diagram* {
        (w) -- [red, line width=0.5mm, -] (m1) -- [red, line width=0.5mm, -] (w0),
    };
    \vertex [circ, below=0em of m1];
    \end{feynman}
    \end{tikzpicture} &  \begin{tikzpicture}[circ/.style={shape=circle, inner sep=2pt, outer sep =0.5pt, line width=0.4mm, draw, node contents=}, baseline={(current bounding box.center)}]
    \begin{feynman}
    \node (m1) ;
    \vertex [left=0.8 em of m1] (w) ;
    \vertex [right=0.8 em of m1] (w0) ;
    \diagram* {
        (w) -- [red, line width=0.5mm, -] (m1) -- [blue, line width=0.5mm, -] (w0),
    };
    \vertex [circ, below=0em of m1];
    \end{feynman}
    \end{tikzpicture} \\[5pt]
 \begin{tikzpicture}[circ/.style={shape=circle, inner sep=2pt, outer sep =0.5pt, line width=0.4mm, draw, node contents=}, baseline={(current bounding box.center)}]
    \begin{feynman}
    \node (m1) ;
    \vertex [left=0.8 em of m1] (w) ;
    \vertex [right=0.8 em of m1] (w0) ;
    \diagram* {
        (w) -- [blue, line width=0.5mm, -] (m1) -- [red, line width=0.5mm, -] (w0),
    };
    \vertex [circ, below=0em of m1];
    \end{feynman}
    \end{tikzpicture} & \begin{tikzpicture}[circ/.style={shape=circle, inner sep=2pt, outer sep =0.5pt, line width=0.4mm, draw, node contents=}, baseline={(current bounding box.center)}]
    \begin{feynman}
    \node (m1) ;
    \vertex [left=0.8 em of m1] (w) ;
    \vertex [right=0.8 em of m1] (w0) ;
    \diagram* {
        (w) -- [blue, line width=0.5mm, -] (m1) -- [blue, line width=0.5mm, -] (w0),
    };
    \vertex [circ, below=0em of m1];
    \end{feynman}
    \end{tikzpicture}\\[3pt]
\end{pmatrix}
\end{equation}
is a transmutation matrix consisting of the four different matrices arising from the integrals over eigenfunctions in Eq.~(\ref{eq:PerturbativeActions}).

For convenience, we define a new matrix
\begin{equation}\label{eq:GeometricMatrixPerturbativeMatrices}
\mathsf{M} \equiv \mathsf{T}\mathsf{G} \equiv
\begin{pmatrix}
\mathsf{A} & \mathsf{B}\\
\mathsf{C} & \mathsf{D}
\end{pmatrix} \mathrel{\hat{=}}
\begin{pmatrix}
\begin{tikzpicture}[circ/.style={shape=circle, inner sep=2pt, outer sep =0.5pt, line width=0.4mm, draw, node contents=}, baseline={(current bounding box.center)}]
    \begin{feynman}
    \node (m1) ;
    \vertex [left=0.8 em of m1] (w) ;
    \vertex [right=2.5 em of m1] (w0) ;
    \diagram* {
        (w) -- [red, line width=0.5mm, -] (m1) -- [red, line width=0.5mm, -] (w0),
    };
    \vertex [circ, below=0em of m1];
    \end{feynman}
    \end{tikzpicture} &  \begin{tikzpicture}[circ/.style={shape=circle, inner sep=2pt, outer sep =0.5pt, line width=0.4mm, draw, node contents=}, baseline={(current bounding box.center)}]
    \begin{feynman}
    \node (m1) ;
    \vertex [left=0.8 em of m1] (w) ;
    \vertex [right=2.5 em of m1] (w0) ;
    \diagram* {
        (w) -- [red, line width=0.5mm, -] (m1) -- [blue, line width=0.5mm, -] (w0),
    };
    \vertex [circ, below=0em of m1];
    \end{feynman}
    \end{tikzpicture} \\[5pt]
 \begin{tikzpicture}[circ/.style={shape=circle, inner sep=2pt, outer sep =0.5pt, line width=0.4mm, draw, node contents=}, baseline={(current bounding box.center)}]
    \begin{feynman}
    \node (m1) ;
    \vertex [left=0.8 em of m1] (w) ;
    \vertex [right=2.5 em of m1] (w0) ;
    \diagram* {
        (w) -- [blue, line width=0.5mm, -] (m1) -- [red, line width=0.5mm, -] (w0),
    };
    \vertex [circ, below=0em of m1];
    \end{feynman}
    \end{tikzpicture} & \begin{tikzpicture}[circ/.style={shape=circle, inner sep=2pt, outer sep =0.5pt, line width=0.4mm, draw, node contents=}, baseline={(current bounding box.center)}]
    \begin{feynman}
    \node (m1) ;
    \vertex [left=0.8 em of m1] (w) ;
    \vertex [right=2.5 em of m1] (w0) ;
    \diagram* {
        (w) -- [blue, line width=0.5mm, -] (m1) -- [blue, line width=0.5mm, -] (w0),
    };
    \vertex [circ, below=0em of m1];
    \end{feynman}
    \end{tikzpicture}\\[3pt]
\end{pmatrix},
\end{equation}
where the explicit forms of the submatrices $\mathsf{A}, \mathsf{B}, \mathsf{C}$ and $\mathsf{D}$ are given in Appendix \ref{app:PerturbativeMatrixAlgebra}.

As the elements of $\mathsf{M}$ are infinite-dimensional matrices, the inverse of $(\mathbb{1}-\mathsf{M})$ in Eq.~(\ref{eq:FeynmanMatrixExpansion}) is not readily calculated. To make progress, we first calculate the blockwise inverse of $(\mathbb{1}-\mathsf{M})$, which gives \footnote{To obtain Eq.~(\ref{eq:BlockwiseInversion}), we define
\begin{equation*}
\begin{split}
    (\mathbb{1}-\mathsf{M})^{-1} &\equiv \begin{pmatrix}
\mathsf{W} & \mathsf{X}\\
\mathsf{Y} & \mathsf{Z}
\end{pmatrix},
\end{split}
\end{equation*}
such that
\begin{equation*}
\begin{split}
    \begin{pmatrix}
\mathbb{1} - \mathsf{A} & \mathsf{B}\\
\mathsf{C} & \mathbb{1} - \mathsf{D}
\end{pmatrix}
\begin{pmatrix}
\mathsf{W} & \mathsf{X}\\
\mathsf{Y} & \mathsf{Z}
\end{pmatrix} &= \begin{pmatrix}
\mathbb{1} & 0\\
0 & \mathbb{1}
\end{pmatrix},\\
\begin{pmatrix}
\mathsf{W} & \mathsf{X}\\
\mathsf{Y} & \mathsf{Z}
\end{pmatrix}
\begin{pmatrix}
\mathbb{1} - \mathsf{A} & \mathsf{B}\\
\mathsf{C} & \mathbb{1} - \mathsf{D}
\end{pmatrix}
&= \begin{pmatrix}
\mathbb{1} & 0\\
0 & \mathbb{1}
\end{pmatrix}.
\end{split}
\end{equation*}
The 8 relations that can be extracted from these expressions, such as $\left(\mathbb{1}-\mathsf{A}\right)\mathsf{W} + \mathsf{B}\mathsf{Y} = \mathbb{1}$, can then be used to solve for the four unknowns $\mathsf{W}$, $\mathsf{X}$, $\mathsf{Y}$ and $\mathsf{Z}$, by writing them in terms of $\mathbb{1}$, $\mathsf{A}$, $\mathsf{B}$, $\mathsf{C}$ and $\mathsf{D}$. The resulting expression for $\left(\mathbb{1}-\mathsf{M}\right)^{-1}$ immediately gives Eq.~(\ref{eq:BlockwiseInversion}).}
\begin{widetext}
\begin{equation}\label{eq:BlockwiseInversion}
\mathsf{F} = \deltabar(\omega + \omega_{0})~\mathsf{G}
\begin{pmatrix}
\left(\mathbb{1} - \mathsf{B}(\mathbb{1}-\mathsf{D})^{-1}\mathsf{C} - \mathsf{A}  \right)^{-1} & 0 \\
0 & \left(\mathbb{1} - \mathsf{C}(\mathbb{1}-\mathsf{A})^{-1}\mathsf{B} - \mathsf{D}  \right)^{-1}
\end{pmatrix}\begin{pmatrix}
\mathbb{1} & \mathsf{B}(\mathbb{1}-\mathsf{D})^{-1} \\
\mathsf{C}(\mathbb{1}-\mathsf{A})^{-1} & \mathbb{1}
\end{pmatrix}.
\end{equation}
\end{widetext}
The full propagators can now be read off from Eq.~(\ref{eq:BlockwiseInversion}), for instance
\begin{equation}\label{eq:FullPropagatorNotSolved}
\begin{tikzpicture}[dot/.style={fill,circle,inner sep=0pt,outer sep=0pt,minimum size=8pt,label={[label distance=0cm]#1}}, baseline={(current bounding box.center)}]
    \begin{feynman}
    \node [dot] (m1) ;
    \vertex [   left=2 em of m1] (w) ;
    \vertex [      right=2 em of m1] (w0) ;
    \diagram* {
        (w) -- [red, line width=0.5mm, -] (m1) -- [red, line width=0.5mm, -] (w0),
    };
    \vertex [below=1.2 em of m1] {\(\)};
    \end{feynman}
    \end{tikzpicture} = \begin{tikzpicture}[dot/.style={fill,circle,inner sep=0pt,outer sep=0pt,minimum size=10pt,label={[label distance=0.3cm]#1}}, baseline={(current bounding box.center)}]
    \begin{feynman}
    \vertex (w) ;
    \vertex [      right=3em of w] (w0) ;
    \diagram* {
        (w) -- [red, line width=0.5mm, -] (w0),
    };
    \vertex [below=0.3 em of m1] {\(\)};
    \end{feynman}
    \end{tikzpicture}
    \left(\mathbb{1} - \mathsf{B}(\mathbb{1}-\mathsf{D})^{-1}\mathsf{C} - \mathsf{A}  \right)^{-1}.
\end{equation}
Obtaining a form for this propagator that can be readily transformed back to real space requires rewriting the remaining inverse matrices on the right-hand side of Eq.~(\ref{eq:FullPropagatorNotSolved}) in closed form. The difficulty involved in calculating the inverse of these infinite-dimensional matrices is avoided by expanding the inverse matrices as a geometric series. The resulting matrix multiplications obey a group-like structure that can be used to reduce the series to a simpler expression. For instance, $\mathsf{D}^{2} = f\mathsf{D}$ where $f = f(\omega)$ is a scalar, see Appendix \ref{app:MatrixMultiplicationEigenvalues}. As a result, $(\mathbb{1}-\mathsf{D})^{-1}$ in Eq.~(\ref{eq:FullPropagatorNotSolved}) reduces to $\mathbb{1}+\mathsf{D}+f\mathsf{D} + \dots = \mathbb{1} + (1-f)^{-1}\mathsf{D}$. Similarly useful identities between the other matrices can be used to remove the remaining inverses in Eq.~(\ref{eq:FullPropagatorNotSolved}). The full details are left to Appendix \ref{app:MathematicalObjects}. Ultimately, Eq.~(\ref{eq:FullPropagatorNotSolved}) reduces to
\begin{equation}\label{eq:FullPropagatorFourier}
\begin{tikzpicture}[dot/.style={fill,circle,inner sep=0pt,outer sep=0pt,minimum size=8pt,label={[label distance=0cm]#1}}, baseline={(current bounding box.center)}]
    \begin{feynman}
    \node [dot] (m1) ;
    \vertex [   left=2 em of m1] (w) ;
    \vertex [      right=2 em of m1] (w0) ;
    \diagram* {
        (w) -- [red, line width=0.5mm, -] (m1) -- [red, line width=0.5mm, -] (w0),
    };
    \vertex [below=1.2 em of m1] {\(\)};
    \end{feynman}
    \end{tikzpicture} = \begin{tikzpicture}[dot/.style={fill,circle,inner sep=0pt,outer sep=0pt,minimum size=10pt,label={[label distance=0.3cm]#1}}, baseline={(current bounding box.center)}]
    \begin{feynman}
    \vertex (w) ;
    \vertex [      right=3em of w] (w0) ;
    \diagram* {
        (w) -- [red, line width=0.5mm, -] (w0),
    };
    \vertex [below=0.3 em of m1] {\(\)};
    \end{feynman}
    \end{tikzpicture}
    \left(\mathbb{1} + \frac{(1-f)~\mathsf{A}+g~\mathsf{A}^{'}}{(1-f)^{2}-g^{2}} \right),
\end{equation}
where the explicit form for the scalar $g = g(\omega)$ is given in Appendix \ref{app:MatrixMultiplicationEigenvalues} and the explicit forms for the matrix $\mathsf{A}^{'}$, and the other primed matrices, are given in Appendix \ref{app:PerturbativeMatrixAlgebra}. Similarly, we find for the propagator representing initialisation of a left-moving particle,
\begin{equation}\label{eq:FullPropagatorFourierLeftMover}
\begin{tikzpicture}[dot/.style={fill,circle,inner sep=0pt,outer sep=0pt,minimum size=8pt,label={[label distance=0cm]#1}}, baseline={(current bounding box.center)}]
    \begin{feynman}
    \node [dot] (m1) ;
    \vertex [   left=2 em of m1] (w) ;
    \vertex [      right=2 em of m1] (w0) ;
    \diagram* {
        (w) -- [red, line width=0.5mm, -] (m1) -- [blue, line width=0.5mm, -] (w0),
    };
    \vertex [below=1.2 em of m1] {\(\)};
    \end{feynman}
    \end{tikzpicture} = \begin{tikzpicture}[dot/.style={fill,circle,inner sep=0pt,outer sep=0pt,minimum size=10pt,label={[label distance=0.3cm]#1}}, baseline={(current bounding box.center)}]
    \begin{feynman}
    \vertex (w) ;
    \vertex [      right=3em of w] (w0) ;
    \diagram* {
        (w) -- [red, line width=0.5mm, -] (w0),
    };
    \vertex [below=0.3 em of m1] {\(\)};
    \end{feynman}
    \end{tikzpicture}
    ~\frac{(1-f)~\mathsf{B}+g~\mathsf{B}^{'}}{(1-f)^{2}-g^{2}}.
\end{equation}

The propagators involving the annihilator field $\psi$ (those with an outgoing blue leg representing a left-moving particle) can be obtained by symmetry from Eqs.~(\ref{eq:FullPropagatorFourier}) and (\ref{eq:FullPropagatorFourierLeftMover}), specifically by exchanging red and blue lines of the Feynman diagrams. For $\langle \psi(\omega) \tilde{\psi}(\omega_{0})\rangle$, this amounts to swapping the leading red line on the right-hand side of Eq.~(\ref{eq:FullPropagatorFourier}) for a blue line and replacing $\mathsf{A}$ and $\mathsf{A}'$ with $\mathsf{D}$ and $\mathsf{D}'$ respectively. Similarly, $\langle\psi(\omega) \tilde{\phi}(\omega_{0})\rangle$ is obtained from Eq.~(\ref{eq:FullPropagatorFourierLeftMover}) by replacing the leading red line with a blue line whilst replacing $\mathsf{B}$ and $\mathsf{B}'$ with $\mathsf{C}$ and $\mathsf{C}'$ respectively.

The behaviour of the propagators in real space and direct time is obtained by evaluating the contour integral enclosing all of their poles along the negative imaginary axis. The characteristic timescale is determined by the pole with the largest non-zero imaginary part. Defining this pole to occur at $\omega = \omega_{c}$, the long-time behaviour of the system $t \rightarrow \infty$ is determined by $P_{R}(x,t) - P_{R}(x) \propto \exp(-\mathring{\imath}\omega_{c}t)$, where $P_{R}(x)$ is the steady-state particle density. The steady-state density is obtained from the residue of the only vanishing pole $\omega = -\mathring{\imath}r \rightarrow 0^{+}$ of the full propagators. The explicit inverse Fourier transform of only one of Eq.~(\ref{eq:FullPropagatorFourier}) or Eq.~(\ref{eq:FullPropagatorFourierLeftMover}) is required to obtain the full steady-state density of right-moving particles since the density is independent of initialisation provided $\alpha>0$ \footnote{The condition $\alpha>0$ is necessary for the steady state to be independent of intialisation by allowing at least some transmutation between species. This can be seen in the invertibility of the propagators: setting $\alpha = 0$ leads to $\langle\phi(\omega) \tilde{\phi}(\omega_{0})\rangle = \langle\phi(\omega) \tilde{\phi}(\omega_{0})\rangle_{0}$ in Eq.~(\ref{eq:FullPropagatorFourier}) and $\langle\phi(\omega) \tilde{\psi}(\omega_{0})\rangle = 0$ in Eq.~(\ref{eq:FullPropagatorFourierLeftMover}). In direct time, the former has the same steady state as Eq.~(\ref{eq:ProbabilityDensityNoTumbling}) but the latter vanishes.}. Hence, the inverse Fourier transform is calculated by either replacing $\langle \phi_{n}(\omega)\tilde{\phi}_{m}(\omega_{0})\rangle_{0}$ in Eq.~(\ref{eq:InverseFourierTransform}) with $\langle \phi_{n}(\omega)\tilde{\phi}_{m}(\omega_{0})\rangle$, Eq.~(\ref{eq:FullPropagatorFourier}), or by replacing $\langle \phi_{n}(\omega)\tilde{\phi}_{m}(\omega_{0})\rangle_{0}$ with $\langle \phi_{n}(\omega)\tilde{\psi}_{m}(\omega_{0})\rangle$, Eq.~(\ref{eq:FullPropagatorFourierLeftMover}), and $\tilde{U}_{m}(x_{0})$ with $\tilde{V}_{m}(x_{0})$. The stationary density that arises from either route is
\begin{equation}\label{eq:ProbabilityDensityNotNeat}
    P_{R}(x) = \frac{1}{2L}\left(\frac{\mathrm{Pe}}{e^{\mathrm{Pe}}-1}e^{\frac{vx}{D}}+e^{\frac{vx}{2D}}C_{\mathrm{Pe}}\frac{2v}{L}h(x)\right),
\end{equation}
where
\begin{equation}\label{eq:InfiniteFourierSeries}
    h(x) = \sum_{n=1}^{\infty}\frac{1-(-1)^{n}e^{-\mathrm{Pe}/2}}{\left(Dk_{n}^{2}+\frac{v^{2}}{4D}\right)\sqrt{1+\frac{v^{2}}{4D^{2}k_{n}^{2}}}}\cos(k_{n}x+\theta_{n}),
\end{equation}
and
\begin{equation}\label{eq:ConstantInProbability}
\begin{split}
    C_{\mathrm{Pe}} &= \frac{\mathrm{Pe}}{e^{-\mathrm{Pe}}+\mathrm{Pe}-1+\frac{v}{\ell\alpha}\left(1-e^{-\mathrm{Pe}}\right)}\\
    &= \frac{\mathrm{Pe}}{e^{-\mathrm{Pe}}+\mathrm{Pe}-1+\frac{v^{2}}{D\alpha}},
    \end{split}
\end{equation}
and the explicit form of $\ell$ in Eq.~(\ref{eq:DiffusionLengthScale}) has been inserted into Eq.~(\ref{eq:ConstantInProbability}) to afford a simpler expression. Eq.~(\ref{eq:InfiniteFourierSeries}) is a series representation of
\begin{equation}\label{eq:ClosedFourierSeries}
    h(x) = \frac{L}{2v}\left(e^{-\frac{vx}{2D}}-\frac{\mathrm{Pe}}{e^{\mathrm{Pe}}-1}e^{\frac{vx}{2D}} \right),
\end{equation}
see Appendix \ref{app:ClosedFormDerivation}. Hence, Eq.~(\ref{eq:ProbabilityDensityNotNeat}) has the closed form
\begin{equation}\label{eq:ProbabilityDensityNeat}
\begin{split}
    P_{R}(x) &= \frac{1}{2L}\left[\left(1-C_{\mathrm{Pe}} \right)\frac{\mathrm{Pe}}{e^{\mathrm{Pe}}-1}e^{\frac{vx}{D}} + C_{\mathrm{Pe}}\right]\\
    &= \frac{1}{2}(1-C_{\mathrm{Pe}})P_{0}(x) + \frac{C_{\mathrm{Pe}}}{2L},
\end{split}
\end{equation}
where the second equality shows $P_{R}(x)$ is, up to a constant shift, proportional to the steady-state density for a drift-diffusive particle without boundary tumbling $P_{0}(x)$, Eq.~(\ref{eq:ProbabilityDensityNoTumbling}). Both the shift and the proportionality factor are essentially determined by $C_{\mathrm{Pe}}$, Eq.~(\ref{eq:ConstantInProbability}), which therefore captures the entire effect of the tumbling.

Eq.~(\ref{eq:ProbabilityDensityNeat}) is the steady-state solution of the boundary tumbling Fokker-Planck equation, Eq.~(\ref{eq:FokkerPlanckCoupledTumbling}), for the density of right-moving particles and is a key result of the present work. The solution for the density of left-moving particles $P_{L}(x)$ can be found immediately by symmetry, since the steady-state density of left-moving particles is a reflection of that of right-moving particles about $x=L/2$. Thus, $P_{L}(x) = P_{R}(L-x)$.

\begin{figure*}
\subfloat[$\mathrm{Pe}=25$, $C_{\mathrm{Pe}} \approx 1.04$, $\ell/L \approx 0.04$\label{subfig:HighPe}]{%
  \includegraphics[width=0.757\columnwidth, trim = 0.5cm 10.6cm 19.2cm 0.2cm, clip]{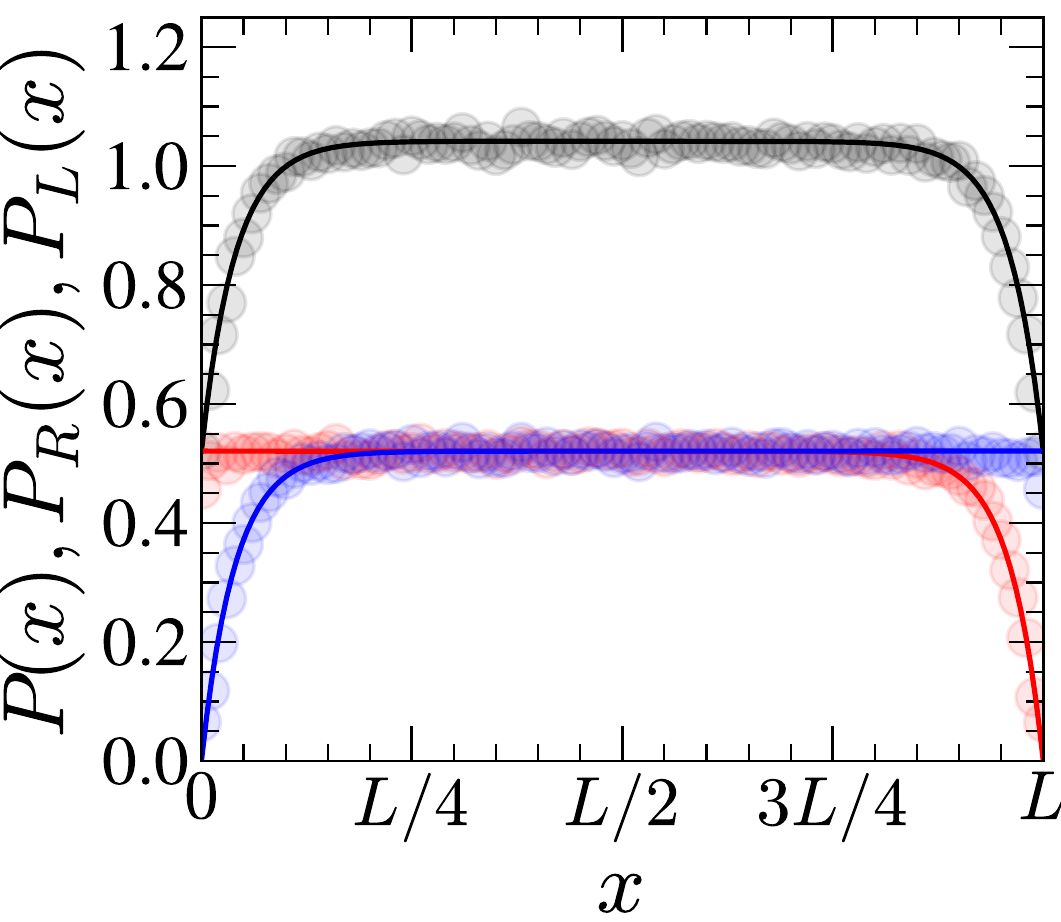}%
}\hfill
\subfloat[$\mathrm{Pe}=\mathrm{Pe}_{\text{min}}\approx 5.54$, $C_{\mathrm{Pe}} \approx 1.22$, $\ell/L \approx 0.18$\label{subfig:MediumPe}]{%
  \includegraphics[width=0.63\columnwidth, trim = 11.7cm 10.6cm 9.8cm 0.2cm, clip]{figures/SimulationComparison.pdf}%
}\hfill
\subfloat[$\mathrm{Pe}=1$, $C_{\mathrm{Pe}} = e \approx 2.72$, $\ell/L \approx 0.63$\label{subfig:LowPe}]{%
  \includegraphics[width=0.63\columnwidth, trim = 21.1cm 10.6cm 0.4cm 0.2cm, clip]{figures/SimulationComparison.pdf}%
}

\subfloat[$\alpha = 100 > \frac{v}{\ell} \approx 30.82$, $C_{\mathrm{Pe}} \approx 1.14$\label{subfig:HighAlpha}]{%
  \includegraphics[width=0.757\columnwidth, trim = 0.5cm 0.6cm 19.2cm 10.1cm, clip]{figures/SimulationComparison.pdf}%
}\hfill
\subfloat[$\alpha = \frac{v}{\ell} \approx 30.82$, $C_{\mathrm{Pe}} = 1$\label{subfig:MediumAlpha}]{%
  \includegraphics[width=0.63\columnwidth, trim = 11.7cm 0.6cm 9.8cm 10.1cm, clip]{figures/SimulationComparison.pdf}%
}\hfill
\subfloat[$\alpha = 10 < \frac{v}{\ell} \approx 30.82$, $C_{\mathrm{Pe}} \approx 0.73$\label{subfig:LowAlpha}]{%
  \includegraphics[width=0.63\columnwidth, trim = 21.1cm 0.6cm 0.4cm 10.1cm, clip]{figures/SimulationComparison.pdf}%
}
\caption{Comparison between the theoretical results (solid lines) for the stationary particle densities, $P_{R}(x)$, Eq.~(\ref{eq:ProbabilityDensityNeat}), $P_{L}(x)$ and $P(x)$, Eq.~(\ref{eq:ProbabilityDensityColourless}), and Monte-Carlo simulations (symbols) for varying $\mathrm{Pe}$, (a)-(c), and varying $\alpha$, (d)-(f). Simulation parameters were set as $D=1$ and $L=1$ such that the self-propulsion speed $v$ fully determines the P\'{e}clet number, i.e. $\mathrm{Pe} = v$. In (a)-(c), $\alpha \rightarrow \infty$ and in (d)-(f), $\mathrm{Pe} = \mathrm{Pe}_{\text{min}} \approx 5.54$, see discussion after Eq.~(\ref{eq:Variance}). The interval is divided into a lattice with spacing $\Delta x = L/100$. In each of the $10^{6}$ simulation runs, a particle is initialised at time $t = 0$ randomly in one of the two internal states, $v$ or $-v$, and at a random site, before proceeding to hop between sites with rates $h_{r} = 2D/\Delta x^{2} + v/\Delta x$ and $h_{l} = 2D/\Delta x^{2} - v/\Delta x$ to the right and to the left respectively \cite{van1992stochastic}. Upon reaching the right-most site, a right-moving particle converts to a left-moving particle with rate $\ell\alpha/\Delta x$, and vice versa for a left-moving particle at the left-most site. This conversion rate results in an effective waiting time $\tau = 1/\alpha$ at the boundaries when diffusive fluctuations are taken into account. Estimates of the density were obtained from frequency histograms of the particle's final position after allowing a suitably long time $t=5$ to elapse for the simulations to converge to the theoretical steady state.}\label{fig:SimulationComparison}
\end{figure*}

A physically relevant observable is the particle density $P(x)$ independent of whether particles are right-moving or left-moving. Since the tumbling is symmetric, a particle in the steady state is right-moving or left-moving with equal probability. Hence,
\begin{equation}\label{eq:ProbabilityDensityColourless}
\begin{split}
    P(x) &\equiv P_{R}(x) + P_{L}(x)\\
    &= \frac{1}{L}\left[\left(1-C_{\mathrm{Pe}} \right)\frac{\mathrm{Pe}~\cosh\left[\mathrm{Pe}\left(\frac{x}{L}-\frac{1}{2}\right)\right]}{2\sinh\left(\frac{\mathrm{Pe}}{2}\right)} + C_{\mathrm{Pe}}\right],
\end{split}
\end{equation}
which has the same functional form as the stationary density for confined RnT particles with uniform tumble rate \cite{frydel2022run}. Eqs.~(\ref{eq:ProbabilityDensityNeat}) and (\ref{eq:ProbabilityDensityColourless}) are in excellent agreement with Monte-Carlo simulations of the boundary tumbling process, Fig.~\ref{fig:SimulationComparison}.

As a further check, we performed simulations to compare the average time $t_{L}$ it takes the particle to complete a length of the system and transmute between species for various $\alpha$. After adjusting for the transmutation, it was found that particles with large $\alpha$ all took the same time $t_{L} - 1/\alpha$ to travel between the boundaries. This confirms $\ell$ in Eq.~(\ref{eq:DiffusionLengthScale}) correctly reproduces the desired waiting time $\tau = 1/\alpha$ for a drift-diffusive particle to transmute at the boundaries. The simulations were performed on a lattice, as opposed to directly simulating the Langevin equation, Eq.~(\ref{eq:Langevin}), to incorporate the reaction probability's dependence on lattice spacing $\Delta x$, Section~\ref{sec:BoundaryTumblingModel}.

As an aside, the transformation $v \rightarrow -v$ immediately obtains the solution to a system where a right-moving particle changes direction only at the left-hand boundary (and vice versa for the left-moving particle).

\subsection{Discussion}\label{sec:BoundaryTumblingDiscussion}

The extent to which the boundary tumbling system's behaviour differs from that of the pure drift-diffusive system without tumbling, Section~\ref{sec:DriftDiffusion}, is captured through $C_{\mathrm{Pe}}$. Vanishing tumble rate $\alpha=0$ results in $C_{\mathrm{Pe}} = 0$, which recovers $P_{R}(x) \propto P_{0}(x)$, Eq.~(\ref{eq:ProbabilityDensityNoTumbling}). For $C_{\mathrm{Pe}} = 1$, the right-moving particle density $P_{R}(x)$ loses all spatial dependence and reduces to a uniform density $P_{R}(x) = 1/2L$. However, $C_{\mathrm{Pe}}>1$ is accessible and in fact guaranteed for $\alpha \rightarrow \infty$,
\begin{equation}\label{eq:CPe_alphaInfinity}
\begin{split}
    \lim_{\alpha\rightarrow\infty}C_{\mathrm{Pe}} = \frac{\mathrm{Pe}}{e^{-\mathrm{Pe}}+\mathrm{Pe}-1} > 1,~\text{for }\mathrm{Pe}>0.
\end{split}
\end{equation}
As seen in Eq.~(\ref{eq:ProbabilityDensityColourless}), $C_{\mathrm{Pe}} > 1$ results in a negative pre-factor of $\cosh\left[\mathrm{Pe}\left(x/L - 1/2 \right) \right]$, resulting in a unimodal distribution of particles that are mostly aggregated at the centre of the system, Figs.~\ref{subfig:HighPe} to \ref{subfig:HighAlpha}. The typical bimodal distribution expected of confined RnT particles \cite{razin2020entropy, malakar2018steady} is obtained for $C_{\mathrm{Pe}} < 1$, Fig.~\ref{subfig:LowAlpha}. Hence, $C_{\mathrm{Pe}} = 1$, Fig.~\ref{subfig:MediumAlpha}, is the point at which the system undergoes a `shape transition' between distributions which resemble active-like, $C_{\mathrm{Pe}} < 1$, and passive-like, $C_{\mathrm{Pe}} > 1$, behaviour \cite{dhar2019run}.

In parameter space, $C_{\mathrm{Pe}} = 1$ corresponds to the curve $\alpha = v / \ell$, Eq.~(\ref{eq:ConstantInProbability}). For $\alpha < v/\ell$ there is a greater tendency for particles to accumulate at the boundaries because of the longer waiting time $\tau = 1/\alpha$ to react, Fig.~\ref{subfig:LowAlpha}. This is otherwise known as the attracting boundaries case, whilst $\alpha > v/\ell$ corresponds to repelling boundaries \cite{angelani2017confined}. Along the curve $\alpha = v / \ell$, the tendency of particles to aggregate at the centre or at the boundaries cancels exactly, which can be seen to result in uniform particle density $P(x) = 1/L$ in Eq.~(\ref{eq:ProbabilityDensityColourless}) and Fig.~\ref{subfig:MediumAlpha}. 

The equality $\alpha = v /\ell$ itself represents a compromise between the time gained by reacting in the diffusive boundary layer of length $\ell$ and the time lost by waiting to react $\tau = 1/\alpha$. Upon reaching the edge of the diffusive boundary layer, the particle is in range of its diffusive fluctuations bringing it in contact with the boundary. Thus the particle \emph{effectively} reacts within the boundary layer of length $\ell$. In the case where $\alpha\rightarrow\infty$, a particle that enters the boundary layer will instantly jump to the boundary and convert to the opposite species. Hence, in this limit, $\ell$ takes on the role of a depletion region length, as can be seen by comparing $\ell/L = (1-\exp(-\mathrm{Pe}))/\mathrm{Pe}$ in Figs.~\ref{subfig:HighPe} to \ref{subfig:LowPe}. This shortens the particle's journey between boundaries by the boundary layer length $\ell$ and hence the travel time shortens on average by $\ell/v$. Uniform density is restored by introducing a finite reaction rate $\alpha = v/\ell$ that exactly compensates for the effective shortening of the system length by making particles wait an average time $\tau = 1/\alpha = \ell/v$ in the boundary layer.
\begin{figure}
        \centering
        \includegraphics[width=0.48\textwidth, trim=1.3cm 0cm 4.7cm 0.8cm, clip]{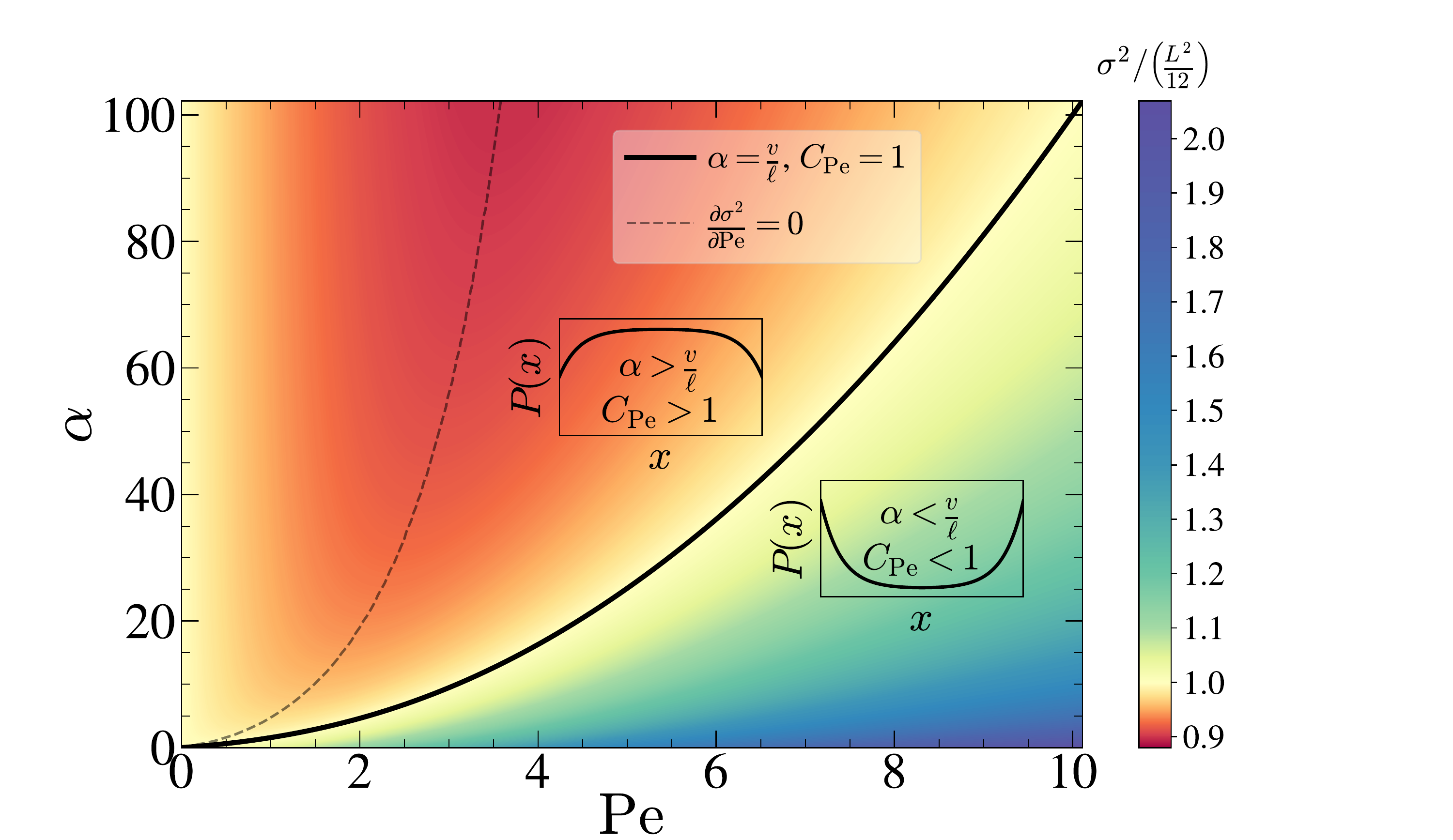}\caption{Variance $\sigma^{2}$, Eq.~(\ref{eq:Variance}), of the particle density $P(x)$, Eq.~(\ref{eq:ProbabilityDensityColourless}), as a function of $\alpha$ and $\mathrm{Pe}$. Parameters were set as $D=1$ and $L=1$ such that $\mathrm{Pe} = v$. The solid line $\alpha = v/\ell$, corresponding to $C_{\mathrm{Pe}} = 1$, divides the plane into two halves. The top half corresponds to unimodal $P(x)$ and thus smaller variance than for a uniform particle density, i.e. $\sigma^{2} < L^{2}/12$. The non-monotonic behaviour of $\sigma^{2}$ as a function of $\mathrm{Pe}$ means there is a curve (dashed line) at which $\sigma^{2}$ is minimised. For bimodal $P(x)$, depicted in the lower half of the plane, $\sigma^{2}>L^{2}/12$ and monotonically increases with $\mathrm{Pe}$.}\label{fig:Variance}
    \end{figure}

A more quantitative way to characterise the shape transition across $C_{\mathrm{Pe}}=1$ is through the variance $\sigma^{2}\equiv \overline{x^{2}} - \overline{x}^{2}$ of the particle density $P(x)$, where $\overline{x^{n}} = \int_{0}^{L}\mathrm{d}x~x^{n}P(x)$ is the $n^{\text{th}}$ moment of $P(x)$. The variance, calculated using $P(x)$ from Eq.~(\ref{eq:ProbabilityDensityColourless}), is
\begin{equation}\label{eq:Variance}
    \sigma^{2} = L^{2}\left[\left(\frac{1}{2} + \frac{2}{\mathrm{Pe}^{2}}-\frac{\coth\left(\frac{\mathrm{Pe}}{2}\right)}{\mathrm{Pe}}\right)\left(1 - C_{\mathrm{Pe}}\right) + \frac{C_{\mathrm{Pe}}}{3} - \frac{1}{4}\right].
\end{equation}
When $\alpha = v/\ell$, corresponding to $C_{\mathrm{Pe}} = 1$, $P(x)$ is uniform and it can be seen that Eq.~(\ref{eq:Variance}) is thus that of a uniform particle density, i.e.\ $\sigma^{2} = L^{2}/12$. When $P(x)$ is unimodal, $\alpha > v/\ell$ and $C_{\mathrm{Pe}} > 1$, particles are aggregated mostly at the centre of the system, meaning they are less spread out, and so $\sigma^{2} < L^{2}/12$. For $\alpha > v/\ell$ there is some non-trivial curve $\mathrm{Pe}_{\text{min}}(\alpha)$ that minimises $\sigma^{2}$ as a function of $\mathrm{Pe}$, Fig.~\ref{fig:Variance}. In principle, this curve can be obtained from $\partial_{\mathrm{Pe}}\sigma^{2} = 0$, but the resulting equation cannot be solved for $\mathrm{Pe}$ in closed form and so it must be solved numerically. Since $\sigma^{2}$ is a monotonically decreasing function of $\alpha$, Fig.~\ref{fig:Variance}, the minimum value of $\sigma^{2}$ occurs on this line at $\alpha\rightarrow\infty$. One finds through numerical evaluation $\mathrm{Pe}_{\text{min}} \equiv \mathrm{Pe}_{\text{min}}(\alpha\rightarrow\infty) \approx 5.54$. Thus $(\mathrm{Pe},\alpha) = (5.54, \infty)$ is the point in parameter space where $\sigma^{2}$ is minimised. The corresponding particle density for $\alpha \rightarrow \infty$ and $\mathrm{Pe} = \mathrm{Pe}_{\mathrm{min}}$ is plotted in Fig.~\ref{subfig:MediumPe}. Figs.~\ref{subfig:HighAlpha} to \ref{subfig:LowAlpha} are also plotted for $\mathrm{Pe} = \mathrm{Pe}_{\mathrm{min}}$. For no activity, $\mathrm{Pe} = 0$, we have that $\sigma^{2} = L^{2}/12$, consistent with the particles being purely diffusive. When the particle density is bimodal, $\alpha < v/\ell$ and $C_{\mathrm{Pe}} < 1$, $\sigma^{2} > L^{2}/12$ since particles are mostly aggregated at the boundaries and are more spread out. In this regime, $\sigma^{2}$ monotonically increases with $\mathrm{Pe}$ as more and more particles accumulate at the boundaries. In the limit $v \rightarrow \infty$, $\sigma^{2} \rightarrow L^{2}/4$ as $P(x) \rightarrow \delta(x)/2 + \delta(x-L)/2$.

We now turn our attention to the limiting behaviour of $P_{R}(x)$, Eq.~(\ref{eq:ProbabilityDensityNeat}). The limiting behaviour of the other particle densities can be obtained immediately from the symmetry arguments $P_{L}(x) = P_{R}(L-x)$ and $P(x) = P_{R}(x) + P_{R}(L-x)$ and so will not be calculated explicitly. In the following, we pay particular attention to whether limits commute with $\alpha \rightarrow \infty$. As we argue below, non-commuting limits reflect that the approach to a particular state can result in different physical features arising in the system.

Firstly, we consider $\mathrm{Pe} \rightarrow 0$. Since $C_{\mathrm{Pe}}$, Eq.~(\ref{eq:ConstantInProbability}), contains another dimensionless parameter, $v^{2}/D\alpha$,  the limits $v \rightarrow 0$ and $D \rightarrow \infty$ are distinguishable and so must be calculated separately. Though right-moving and left-moving particles are physically indistinguishable in the limit $v \rightarrow 0$, we assume they can be identified by some other attribute (such as colour as in Fig.~\ref{fig:Schematic}). Taking the $v \rightarrow 0$ limit first, we have for $P_{R}(x)$, 
\begin{equation}\label{eq:PR_v0_limit}
    \lim_{v \rightarrow 0}P_{R}(x) = \frac{1}{L}\frac{D + \alpha L (L-x)}{2D + \alpha L^{2}}.
\end{equation}
For $\alpha \rightarrow 0$ or $D \rightarrow \infty$ in Eq.~(\ref{eq:PR_v0_limit}), $P_{R}(x) \rightarrow 1/2L$ as in the equilibrium case. For $\alpha \rightarrow \infty$ or $D \rightarrow 0$ in Eq.~(\ref{eq:PR_v0_limit}), right-moving particles deplete linearly between the two boundaries, i.e.\ $P_{R}(x) \rightarrow (1-x/L)/L$, as can be seen for small $\mathrm{Pe}$ in Fig.~\ref{subfig:LowPe}. 

The limit $D\rightarrow \infty$ in Eq.~(\ref{eq:ProbabilityDensityNeat}) is the first instance of non-commutativity with $\alpha \rightarrow \infty$. If $D\rightarrow \infty$ is taken first, then $P_{R}(x) \rightarrow 1/2L$, again in line with an equilibrium system. However, for $\alpha \rightarrow \infty$, the dimensionless parameter $v^{2}/D\alpha$ in $C_{\mathrm{Pe}}$, Eq.~(\ref{eq:ConstantInProbability}), vanishes and so one can simply consider $\mathrm{Pe}\rightarrow 0$ in $C_{\mathrm{Pe}}$, Eq.~(\ref{eq:CPe_alphaInfinity}). Thus, first taking $\alpha \rightarrow \infty$ \emph{then} $D\rightarrow \infty$ gives the same as the $v \rightarrow 0$ limit above, i.e.\ $P_{R}(x) \rightarrow (1-x/L)/L$. Despite these limits not commuting, they retain the same particle density $P(x) \rightarrow 1/L$ as $D\rightarrow \infty$ and $\alpha \rightarrow \infty$.

We now consider the $\mathrm{Pe} \rightarrow \infty$ limits starting with $v \rightarrow \infty$. Treating the boundary as a separate case, we determine $\lim_{v\rightarrow\infty}P_{R}(x < L) = 0$ and $\lim_{v\rightarrow\infty}P_{R}(x=L) = \infty$. We thus assign
\begin{equation}\label{eq:PR_vInfty_limit}
    \lim_{v \rightarrow \infty}P_{R}(x) = \frac{1}{2}\delta(x-L),
\end{equation}
for a correctly normalised density of right-moving particles. Eq.~(\ref{eq:PR_vInfty_limit}) reflects that particles with infinite self-propulsion speed $v$ are not found in the bulk since they immediately jump to the right-hand boundary where they wait to transmute for a finite time $\tau = 1/\alpha$.

Similar care for boundary cases needs to be taken for $D \rightarrow 0$ in Eq.~(\ref{eq:ProbabilityDensityNeat}), for which we find $\lim_{D\rightarrow 0} P_{R}(x<L) = (1+v/L\alpha)^{-1}/2L$ and $\lim_{D\rightarrow 0} P_{R}(x=L) = \mathrm{Pe}[1-(1+v/L\alpha)^{-1}]/2L$, i.e. $P_{R}(x)$ diverges at $x=L$, but is finite in the bulk. Hence,
\begin{equation}\label{eq:PR_D0_limit}
    \lim_{D \rightarrow 0}P_{R}(x) = \frac{1}{2L}\frac{1}{1+\frac{v}{L\alpha}} + \frac{1}{2}\delta(x-L)\left(1 - \frac{1}{1+\frac{v}{L\alpha}}\right).
\end{equation}
Eq.~(\ref{eq:PR_D0_limit}) differs to Eq.~(\ref{eq:PR_vInfty_limit}) because particles still have finite self-propulsion speed $v$ and thus still spend a finite time travelling in the bulk. Taking, $v \rightarrow \infty$ in Eq.~(\ref{eq:PR_D0_limit}) recovers Eq.~(\ref{eq:PR_vInfty_limit}). Taking $\alpha \rightarrow 0$ in Eq.~(\ref{eq:PR_D0_limit}) also gives $P_{R}(x) \rightarrow \delta(x-L)/2$ since particles cannot change direction at the right-hand boundary and so get stuck there. For $\alpha \rightarrow \infty$ in Eq.~(\ref{eq:PR_D0_limit}), we have $P_{R}(x) \rightarrow 1/2L$ since particles never wait to react at the boundary and so are uniformly distributed.

We obtain different limiting behaviour if we take $\alpha \rightarrow \infty$ before the above $\mathrm{Pe}\rightarrow\infty$ limits, i.e.\ $v \rightarrow \infty$ or $D \rightarrow 0$. As before, first taking $\alpha \rightarrow \infty$ restores the symmetry between $v \rightarrow \infty$ and $D \rightarrow 0$ as $\mathrm{Pe}$ becomes the only dimensionless parameter in $C_{\mathrm{Pe}}$, Eq.~(\ref{eq:CPe_alphaInfinity}). Thus, taking either $v \rightarrow \infty$ or $D \rightarrow 0$ \emph{after} $\alpha \rightarrow \infty$ results in
\begin{equation}\label{eq:LimitInfinitePeclet}
   \lim_{\mathrm{Pe} \rightarrow \infty}\lim_{\alpha \rightarrow \infty} P_{R}(x) = \begin{cases}
			\frac{1}{2L}, & 0 \leq x < L\\
            0, & x = L
		 \end{cases},
\end{equation}
reflecting a right-moving particle can never be found at the right-hand boundary where it transmutes instantly, but it should otherwise be uniformly distributed as it is not sensitive to any waiting time at the boundary. This is seen for large $\mathrm{Pe}$ in Fig.~\ref{subfig:HighPe}, where $P_{R}(x)$ approaches uniform density but still has $P_{R}(L) = 0$.

Physically, the non-commuting limits $\alpha \rightarrow \infty$ and $\mathrm{Pe} \rightarrow \infty$ demonstrate that different behaviour is obtained depending on whether $C_{\mathrm{Pe}}$ approaches $1$ from above or below, $C_{\mathrm{Pe}} \rightarrow 1^{\pm}$, in Eq.~(\ref{eq:ProbabilityDensityNeat}). For instance, consider if the system first starts in the state where $\alpha \rightarrow \infty$ and $D,v$ are both finite. This places it in the top half of the plane in Fig.~\ref{fig:Variance}, above the curve $C_{\mathrm{Pe}} = 1$, such that $P(x)$ is unimodal and resembles $P(x)$ in Fig.~\ref{subfig:HighPe}. Then, evolving the system parameters so that $\mathrm{Pe} \rightarrow \infty$, such as by cooling the surrounding particle bath so that $D \rightarrow 0$, shifts the system state in parameter space to the right till it meets the curve $C_{\mathrm{Pe}} = 1$ from above, $C_{\mathrm{Pe}} \rightarrow 1^{+}$. While the system is being evolved, $P_{R}(x)$ is restricted to $P_{R}(L) = 0$ and so retains this feature as the system approaches $C_{\mathrm{Pe}} \rightarrow 1^{+}$, resulting in Eq.~(\ref{eq:LimitInfinitePeclet}). A similar argument applies to first taking $v \rightarrow \infty$ or $D \rightarrow 0$ followed by $\alpha \rightarrow \infty$, which places the system state in the lower half of the plane in Fig.~\ref{fig:Variance} such that $C_{\mathrm{Pe}} \rightarrow 1^{-}$. In this case, the right-moving particle density $P_{R}(x)$ resembles that in Fig.~\ref{subfig:LowAlpha} and so does not have the feature $P_{R}(L) = 0$ as the system is evolved towards $C_{\mathrm{Pe}} = 1$. Hence, the different approaches to $C_{\mathrm{Pe}} = 1$ result in different particle densities in this limit.

To conclude the limiting behaviour of $P_{R}(x)$, for $L \rightarrow 0$, Eq.~(\ref{eq:ProbabilityDensityNeat}) diverges since all particles are confined to a single point in space. In the limit of infinite system size $L \rightarrow \infty$, $P_{R}(x) \rightarrow 0$ because the particle is not confined and so does not have a non-trivial steady state.

We now consider how the current behaves under these same limits. The steady-state current for right-moving particles is given by
\begin{equation}\label{eq:RightMoverCurrent}
   J_{R}(x) \equiv vP_{R}(x) - D\frac{\partial P_{R}(x)}{\partial x} = \frac{v}{2L}C_{\mathrm{Pe}}.
\end{equation}
The current for left-moving particles $J_{L}(x)$ is found from replacing $P_{R}(x)$ with $P_{L}(x)$ and $v$ with $-v$ on the right-hand side of $\equiv$ in Eq.~(\ref{eq:RightMoverCurrent}). This results in $J_{L} = -J_{R}$ and so both currents are constant across space, whilst the overall particle density current $J = J_{R} + J_{L}$ vanishes. The non-zero currents that arise when the internal state is known are a result of a given particle species appearing to continually `enter' the system at one boundary and then `leave' through the other. This is only possible if the particle can transmute between species and thus $\alpha > 0$ is crucial for obtaining perpetual currents in the steady state.

We find that all limits of the current commute with $\alpha \rightarrow \infty$. First, taking $v \rightarrow 0$,
\begin{equation}\label{eq:CurrentV0Limit}
   \lim_{v \rightarrow 0}J_{R} = \frac{\alpha D}{2D + \alpha L^{2}},
\end{equation}
which vanishes for $\alpha = 0$ (no transmutation to generate a steady-state current) or $D = 0$ (particles do not move). Though the particle is no longer active for $v \rightarrow 0$, the transmutation effectively rectifies the motion of a single particle species and so we still observe a current if the species can be distinguished. Taking instead $v \rightarrow \infty$ or $D \rightarrow \infty$ in Eq.~(\ref{eq:RightMoverCurrent}) gives $J_{R} \rightarrow \alpha/2$ since particles traverse the system instantly and so the dynamics effectively reduce to a single point where a particle transmutes between species with rate $\alpha$ (the factor $1/2$ accounts for the probability to find the particle in the right-moving state). This is confirmed by taking $L \rightarrow 0$ in Eq.~(\ref{eq:RightMoverCurrent}), which gives the same result $J_{R} \rightarrow \alpha/2$.

In the deterministic limit $D \rightarrow 0$, Eq.~(\ref{eq:RightMoverCurrent}) becomes \begin{equation}\label{eq:CurrentD0Limit}
   \lim_{D \rightarrow 0}J_{R} = \frac{\frac{1}{2}}{\frac{L}{v} + \frac{1}{\alpha}},
\end{equation}
reflecting a particle takes time $t_{L} = L/v + 1/\alpha$ to complete a length of the system since $L/v$ time is spent moving between the boundaries and $1/\alpha$ time is spent waiting to react at the boundaries. Finally, taking $L \rightarrow \infty$ in Eq.~(\ref{eq:RightMoverCurrent}) gives $J_{R} \rightarrow 0$ due to the lack of a non-trivial steady-state density.

\section{Conclusion}

We have derived the fully time-dependent, Eqs.~(\ref{eq:FullPropagatorFourier}) and (\ref{eq:FullPropagatorFourierLeftMover}), and steady-state, Eqs.~(\ref{eq:ProbabilityDensityNeat}) and (\ref{eq:ProbabilityDensityColourless}), solutions to the particle density of a confined self-propelled particle that changes self-propulsion direction through interacting stochastically with boundaries. The results of this `boundary tumbling' RnT model were confirmed by lattice simulations, Fig.~\ref{fig:SimulationComparison}. We were able to handle localised reactions within a perturbative framework, Section~\ref{sec:BoundaryTumblingSolution}, which posed a significant technical challenge compared to a confined RnT system with uniform tumble rate \cite{razin2020entropy, malakar2018steady}. The derivation made use of field-theoretic techniques in the Doi-Peliti formalism \cite{doi1976second, peliti1985path}, thus making the developed framework easily extensible to including additional arbitrary interactions.

Of particular interest in this work is the incorporation of diffusion, which prevents particles from residing at the boundaries even when their self-propulsion drives them towards the boundaries. While a Poissonian transmutation at the boundaries is trivial to implement and handle on a lattice, where particles can appear to be `stuck' at a wall, doing the same in the continuum posed physical and mathematical hurdles which we addressed in Section~\ref{sec:BoundaryTumblingModel} and Appendix \ref{app:LengthScaleDerivation}.

The results show that boundary RnT particles undergo a shape transition that depends on the size $\ell$, Eq.~(\ref{eq:DiffusionLengthScale}), of the diffusive fluctuations away from the boundaries. This transition is between particles mainly aggregating at the system centre for large transmutation rates $\alpha > v/\ell$ or at the boundaries for small transmutation rates $\alpha < v/\ell$, Figs.~\ref{subfig:HighAlpha} to \ref{subfig:LowAlpha}. This transition is characterised by whether the variance $\sigma^{2}$, Eq.~(\ref{eq:Variance}), of the overall particle density $P(x)$, Eq.~(\ref{eq:ProbabilityDensityColourless}), is greater or smaller than that of a uniform particle density $L^{2}/12$.

The perturbative framework, used to derive the particle density, built on the solution of the reflecting boundaries drift-diffusion Fokker-Planck equation, Eq.~(\ref{eq:BarePropagator}). The same approach could be taken for the study of confined RnT particles with more general non-uniform tumble rates that include a separate boundary tumbling term. This would generalise the results of Refs.~\cite{malakar2018steady, razin2020entropy} to non-uniform tumble rates, and Ref.~\cite{angelani2017confined} to non-vanishing passive diffusion. Applying the scheme developed in the present work to derive the fully time-dependent particle density of such a model is an avenue for further research.

\begin{acknowledgements}
We are grateful to Luca Cocconi, Jacob Knight, Marius Bothe, Letian Chen, Farid Kaveh, Ziluo Zhang, Zigan Zhen and Rosalba Garcia-Millan for useful discussions. This work was supported by the Engineering and Physical Sciences Research Council [grant number EP/V520238/1].
\end{acknowledgements}

\appendix

\section{\label{app:Eigenstates}Right eigenfunctions $U_{n}(x)$ of the reflecting boundaries drift-diffusion equation}

A similar derivation to the following appears in Ref.~\cite{zhang2009comparison}.

The right eigenfunctions of the drift-diffusion equation with reflecting boundaries are derived in general by trialling
\begin{equation}\label{eq:TrialDriftDiffusionEqn}
    U_{n}(x) = A(a)~e^{ax}.
\end{equation}
Substituting Eq.~(\ref{eq:TrialDriftDiffusionEqn}) into the eigenvalue equation
\begin{equation}\label{eq:DriftDiffusionEigenvalueEqn}
    \mathcal{L}U_{n}(x) = \left(D\frac{\partial^{2}}{\partial x^2}- v\frac{\partial}{\partial x}\right)U_{n}(x) = \lambda_{n}U_{n}(x)
\end{equation}
results in the characteristic equation
\begin{equation}\label{eq:DriftDiffusionEigenvalueEqn2}
    Da^{2}-va-\lambda_{n} = 0.
\end{equation}
Solving for the exponent, we find 
\begin{equation}\label{eq:DriftDiffusionSimultaneousSatisfiedEqn}
   a_{\pm} = \frac{v}{2D} \pm \sqrt{\frac{v^{2}}{4D^{2}}+\frac{\lambda_{n}}{D}}.
\end{equation}
Hence, the general solution is the sum of these two solutions, i.e.
\begin{equation}\label{eq:TrialDriftDiffusionWithExponentsEqn}
    U_{n}(x) = A_{+}e^{a_{+}x} + A_{-}e^{a_{-}x}.
\end{equation}

The eigenvalues $\lambda_{n}$ and one of the amplitudes $A_{\pm}$ are fixed by the boundary conditions, as we demonstrate in the following. First, from substituting Eq.~(\ref{eq:TrialDriftDiffusionWithExponentsEqn}) into the boundary condition $J(x=0,t)=0$, Eq.~(\ref{eq:CurrentNoTumbling}),
\begin{equation}\label{eq:BC1}
    \left(D\frac{\partial}{\partial x} - v \right)U_{n}(x)\Big|_{x=0} = 0,
\end{equation}
it follows that
\begin{equation}\label{eq:WaveAmplitudeCondition}
    A_{-} = A_{+} \frac{\sqrt{v^{2}+4\lambda_{n}D}-v}{\sqrt{v^{2}+4\lambda_{n}D}+v}.
\end{equation}
Similarly, applying the boundary condition at $x=L$,
\begin{equation}\label{eq:EigenvaluesCondition}
   \left(\sqrt{\frac{v^{2}}{4}+D\lambda_{n}}-\frac{v}{2} \right)\sinh\left(\sqrt{\frac{v^{2}}{4D^{2}}+\frac{\lambda_{n}}{D}}L\right)=0.
\end{equation}
This implies the eigenvalues are given by $\lambda_{0} = 0$ (from the first bracket) and $\lambda_{n} = -Dk_{n}^{2} - v^{2}/4D$ (from the $\sinh$ function), as stated around Eqs.~(\ref{eq:StationaryRightEigenstateNoTumbling}) and (\ref{eq:RightEigenstatesNoTumbling}).

Beginning with the eigenvalue for the stationary eigenfunction $\lambda_{0} = 0$, Eq.~(\ref{eq:WaveAmplitudeCondition}) gives $A_{-}=0$ and the exponent for the first term in Eq.~(\ref{eq:TrialDriftDiffusionWithExponentsEqn}) reduces to $a_{+}x = vx/D$, recovering the stationary eigenfunction $U_{0}(x) = \exp(vx/D)$, Eq.~(\ref{eq:StationaryRightEigenstateNoTumbling}).

Finally, we consider the non-stationary eigenfunctions with eigenvalues $\lambda_{n} = -Dk_{n}^{2} - v^{2}/4D$. Substituting this into Eq.~(\ref{eq:TrialDriftDiffusionWithExponentsEqn}), we find
\begin{equation}\label{eq:SolvedExpForm}
    U_{n}(x) = A_{+}e^{\frac{v}{2D}x}\left[e^{ik_{n}x} + \frac{2iDk_{n}-v}{2iDk_{n}+v}e^{-ik_{n}x}\right],
\end{equation}
which reduces to Eq.~(\ref{eq:RightEigenstatesNoTumbling}) after rewriting in terms of trigonometric functions and with suitable normalisation determining $A_{+}$.

\section{\label{app:LengthScaleDerivation}Derivation of boundary tumbling length scale $\ell$}

The boundary tumbling length scale $\ell$, Eq.~(\ref{eq:DiffusionLengthScale}), was introduced in Section~\ref{sec:BoundaryTumblingModel} on dimensional grounds and as a way to account for fluctuations keeping a particle away from a boundary where it would otherwise transmute into the opposite species. In the present section, we derive this length scale from a lattice model. We will demand the effective rate of transmutation is $\alpha$ for a particle that appears at the boundary site $i=N$ with a probability $p_N$ according to the steady state of a non-transmuting drift-diffusion particle. We will find the probability to reside at the boundary site decreases with the lattice spacing $\Delta x$ due to fluctuations, reducing the chances for the transmutation to take place at that site. To maintain a constant \emph{effective} transmutation rate $\alpha$, despite diminished residence probability, we will need to compensate for the decreasing residence probability at the transmutation site $i=N$ by increasing the rate with which transmutation takes place conditional to residence. We will consider here only a right-moving particle, because the arguments readily translate to left-moving particles.

Denoting by $p_i$ the steady-state probability to be located at site $i\in\{1,2,\ldots,N\}$ in a system with $N$ sites, a right-moving particle therefore needs to transmute with effective rate $\alpha/p_N$ at the right-most site, captured in a master equation by adding a transmutation term with rate $\delta_{i,N}\alpha/p_N$. If the steady state were such that the particle resides at site $i=N$ almost surely, i.e.\ $p_{N}=1$, the effective transmutation rate would simply be $\alpha$ and the dynamics a matter of the particle reaching the steady state and transmuting over and over. However, if the hopping rates are $h_{r} = 2D/\Delta x^{2} + v/\Delta x$ and $h_{l} = 2D/\Delta x^{2} - v/\Delta x$ to the right and to the left respectively \cite{van1992stochastic}, the root of the steady-state equations $0 = h_{r} p_{i-1}-(h_{r}+h_{l})p_{i}+h_{l}p_{i+1}$ in the bulk and similarly at the boundary sites is
\begin{equation}
p_i=\mathcal{N}^{-1} \left(\frac{h_{r}}{h_{l}}\right)^{i},
\end{equation}
with normalisation $\mathcal{N}=q(1-q^{N})/(1-q)$, where
\begin{equation}
q \equiv \frac{h_{r}}{h_{l}}=1+\frac{v\Delta x}{D} + \mathcal{O}(\Delta x^{2}) \ .
\end{equation}
After some algebra, it follows that $p_{N}$ vanishes like $\Delta x$ and taking the limit $\Delta x \rightarrow 0$ with $N\Delta x=L$ remaining constant gives
\begin{equation}
\lim_{\Delta x \rightarrow 0} \frac{p_{N}}{\Delta x} = \frac{v}{D}\left(1-e^{-\frac{vL}{D}}\right)^{-1}
\ .
\end{equation}
In a master equation, the transmutation therefore needs to take place with rate
\begin{equation}\label{transmutation_continuum_limit}
\begin{split}
\lim_{\Delta x\rightarrow 0} \frac{\delta_{i,N}}{p_{N}}\alpha &= \frac{D}{v} \left(1-e^{-\frac{vL}{D}}\right)\alpha \delta(x-L)\\
&= \ell \alpha \delta(x-L)
\end{split}
\end{equation}
using $\lim_{\Delta x\rightarrow 0}\delta_{i,N}/\Delta x = \delta(x-L)$ and the definition of the length scale $\ell$ of Eq.~(\ref{eq:DiffusionLengthScale}).

Eq.~(\ref{transmutation_continuum_limit}) remains valid when $v<0$, as $\ell \geq 0$, which means particles have a bias towards the boundary where they came into existence by transmutation.

\begin{widetext}
\section{\label{app:MathematicalObjects}Explicit expressions for mathematical objects in Section~\ref{sec:BoundaryTumblingSolution}}

This appendix contains explicit expressions for some of the mathematical objects encountered in Section~\ref{sec:BoundaryTumblingSolution}, namely the integrals over eigenfunctions arising from Eq.~(\ref{eq:PerturbativeActions}), the perturbative matrices $\mathsf{A}$, $\mathsf{B}$, $\mathsf{C}$ and $\mathsf{D}$ defined in Eq.~(\ref{eq:GeometricMatrixPerturbativeMatrices}), and their counterparts $\mathsf{A}'$, $\mathsf{B}'$, $\mathsf{C}'$ and $\mathsf{D}'$ that appear in Eqs.~(\ref{eq:FullPropagatorFourier}) and (\ref{eq:FullPropagatorFourierLeftMover}).

\subsection{\label{app:MatrixIntegrals}Integrals over eigenfunctions $\Delta$}

The integrals over eigenfunctions arising from the perturbative parts of the action, Eq.~(\ref{eq:PerturbativeActions}), are
\begin{equation}\label{eq:PhiPhiIntegral}
\begin{split}
\Delta_{\tilde{\phi}_{i}\phi_{j}} &= -\frac{\ell\alpha}{L^{2}} \int_{0}^{L}\mathrm{d}x~\delta(x-L)~\tilde{U}_{i}(x)U_{j}(x)\\
&= -\frac{\ell\alpha}{L^{2}}\Bigg( \frac{\mathrm{Pe}}{1-e^{-\mathrm{Pe}}}\delta_{i0}\delta_{j0} +\frac{2\mathrm{Pe}}{1-e^{-\mathrm{Pe}}}e^{-\mathrm{Pe}/2}\frac{(-1)^{j}}{\sqrt{1+v^{2}/4D^{2}k_{j}^{2}}}\delta_{i0}(1-\delta_{j0})\\
&~+ 2e^{\mathrm{Pe}/2}\frac{(-1)^{i}}{\sqrt{1+v^{2}/4D^{2}k_{i}^{2}}}(1-\delta_{i0})\delta_{j0} + \frac{4(-1)^{i+j}}{\sqrt{\left(1+v^{2}/4D^{2}k_{i}^{2}\right)\left(1+v^{2}/4D^{2}k_{j}^{2}\right)}}(1-\delta_{i0})(1-\delta_{j0}) \Bigg),
\end{split}
\end{equation}
\begin{equation}\label{eq:PsiPhiIntegral}
\begin{split}
\Delta_{\tilde{\psi}_{i}\phi_{j}} &= \frac{\ell\alpha}{L^{2}} \int_{0}^{L}\mathrm{d}x~\delta(x-L)~\tilde{V}_{i}(x)U_{j}(x)\\
&= -\Delta_{\tilde{\phi}_{i}\phi_{j}}e^{\mathrm{Pe}},
\end{split}
\end{equation}
\begin{equation}\label{eq:PsiPsiIntegral}
\begin{split}
\Delta_{\tilde{\psi}_{i}\psi_{j}} &= -\frac{\ell\alpha}{L^{2}} \int_{0}^{L}\mathrm{d}x~\delta(x)~\tilde{V}_{i}(x)V_{j}(x)\\
&= -\frac{\ell\alpha}{L^{2}}\Bigg( \frac{\mathrm{Pe}}{1-e^{-\mathrm{Pe}}}\delta_{i0}\delta_{j0} +  \frac{2\mathrm{Pe}}{1-e^{-\mathrm{Pe}}}\frac{1}{\sqrt{1+v^{2}/4D^{2}k_{j}^{2}}}\delta_{i0}(1-\delta_{j0})\\
&~+ \frac{2}{\sqrt{1+v^{2}/4D^{2}k_{i}^{2}}}(1-\delta_{i0})\delta_{j0} + \frac{4}{\sqrt{\left(1+v^{2}/4D^{2}k_{i}^{2}\right)\left(1+v^{2}/4D^{2}k_{j}^{2}\right)}}(1-\delta_{i0})(1-\delta_{j0}) \Bigg),
\end{split}
\end{equation}
\begin{equation}\label{eq:PhiPsiIntegral}
\begin{split}
\Delta_{\tilde{\phi}_{i}\psi_{j}} &= \frac{\ell\alpha}{L^{2}} \int_{0}^{L}\mathrm{d}x~\delta(x)~\tilde{U}_{i}(x)V_{j}(x)\\
&= \frac{\ell\alpha}{L^{2}}\Bigg( \frac{\mathrm{Pe}}{1-e^{-\mathrm{Pe}}}e^{-\mathrm{Pe}}\delta_{i0}\delta_{j0} +  \frac{2\mathrm{Pe}}{1-e^{-\mathrm{Pe}}}e^{-\mathrm{Pe}}\frac{1}{\sqrt{1+v^{2}/4D^{2}k_{j}^{2}}}\delta_{i0}(1-\delta_{j0})\\
&~+ \frac{2}{\sqrt{1+v^{2}/4D^{2}k_{i}^{2}}}(1-\delta_{i0})\delta_{j0} + \frac{4}{\sqrt{\left(1+v^{2}/4D^{2}k_{i}^{2}\right)\left(1+v^{2}/4D^{2}k_{j}^{2}\right)}}(1-\delta_{i0})(1-\delta_{j0}) \Bigg).
\end{split}
\end{equation}

\subsection{\label{app:PerturbativeMatrixAlgebra}Perturbative matrices $\mathsf{A}$, $\mathsf{A}'$ etc.}

The perturbative matrices $\mathsf{A}$, $\mathsf{B}$, $\mathsf{C}$ and $\mathsf{D}$, Eq.~(\ref{eq:GeometricMatrixPerturbativeMatrices}), are calculated from the integrals, Eqs.~(\ref{eq:PhiPhiIntegral}) to (\ref{eq:PhiPsiIntegral}), and the bare propagator, Eq.~(\ref{eq:BarePropagator}), via $\mathsf{A}_{ij} = \sum_{k} \Delta_{\tilde{\phi}_{i}\phi_{k}}\langle\phi_{k}(\omega)\tilde{\phi}_{j}(\omega_{0})\rangle$ and similar for $\mathsf{B}$, $\mathsf{C}$ and $\mathsf{D}$. The primed counterparts to these matrices $\mathsf{A}'$, $\mathsf{B}'$, $\mathsf{C}'$ and $\mathsf{D}'$, that appear in Eqs.~(\ref{eq:FullPropagatorFourier}) and (\ref{eq:FullPropagatorFourierLeftMover}), arise as products of the unprimed matrices, Appendix \ref{app:MatrixMultiplicationEigenvalues}. The explicit forms of these matrices are
\begin{equation}\label{eq:MatrixA}
\begin{split}
\mathsf{A}_{ij} = -\mathsf{C}_{ij}e^{-\mathrm{Pe}} &= -\frac{\ell\alpha}{L}\Bigg( \frac{\mathrm{Pe}}{1-e^{-\mathrm{Pe}}}\frac{\delta_{i0}\delta_{j0}}{-\mathring{\imath}\omega + r} +  \frac{\mathrm{Pe}}{1-e^{-\mathrm{Pe}}}e^{-\mathrm{Pe}/2}\frac{(-1)^{j}}{\sqrt{1+v^{2}/4D^{2}k_{j}^{2}}}\frac{\delta_{i0}(1-\delta_{j0})}{-\mathring{\imath}\omega + Dk_{j}^{2} + \frac{v^{2}}{4D} + r}\\
&~+ 2e^{\mathrm{Pe}/2}\frac{(-1)^{i}}{\sqrt{1+v^{2}/4D^{2}k_{i}^{2}}}\frac{(1-\delta_{i0})\delta_{j0}}{-\mathring{\imath}\omega + r} + \frac{2(-1)^{i+j}}{\sqrt{\left(1+v^{2}/4D^{2}k_{i}^{2}\right)\left(1+v^{2}/4D^{2}k_{j}^{2}\right)}}\frac{(1-\delta_{i0})(1-\delta_{j0})}{-\mathring{\imath}\omega + Dk_{j}^{2} + \frac{v^{2}}{4D} + r} \Bigg),
\end{split}
\end{equation}
\begin{equation}\label{eq:MatrixD}
\begin{split}
\mathsf{D}_{ij} &= -\frac{\ell\alpha}{L}\Bigg( \frac{\mathrm{Pe}}{1-e^{-\mathrm{Pe}}}\frac{\delta_{i0}\delta_{j0}}{-\mathring{\imath}\omega + r} +  \frac{\mathrm{Pe}}{1-e^{-\mathrm{Pe}}}\frac{1}{\sqrt{1+v^{2}/4D^{2}k_{j}^{2}}}\frac{\delta_{i0}(1-\delta_{j0})}{-\mathring{\imath}\omega + Dk_{j}^{2} + \frac{v^{2}}{4D} + r}\\
&~+ \frac{2}{\sqrt{1+v^{2}/4D^{2}k_{i}^{2}}}\frac{(1-\delta_{i0})\delta_{j0}}{-\mathring{\imath}\omega + r} + \frac{2}{\sqrt{\left(1+v^{2}/4D^{2}k_{i}^{2}\right)\left(1+v^{2}/4D^{2}k_{j}^{2}\right)}}\frac{(1-\delta_{i0})(1-\delta_{j0})}{-\mathring{\imath}\omega + Dk_{j}^{2} + \frac{v^{2}}{4D} + r} \Bigg),
\end{split}
\end{equation}
\begin{equation}\label{eq:MatrixB}
\begin{split}
\mathsf{B}_{ij} &= \frac{\ell\alpha}{L}\Bigg( \frac{\mathrm{Pe}}{1-e^{-\mathrm{Pe}}}e^{-\mathrm{Pe}}\frac{\delta_{i0}\delta_{j0}}{-\mathring{\imath}\omega + r} +  \frac{\mathrm{Pe}}{1-e^{-\mathrm{Pe}}}\frac{e^{-\mathrm{Pe}}}{\sqrt{1+v^{2}/4D^{2}k_{j}^{2}}}\frac{\delta_{i0}(1-\delta_{j0})}{-\mathring{\imath}\omega + Dk_{j}^{2} + \frac{v^{2}}{4D} + r}\\
&~+ \frac{2}{\sqrt{1+v^{2}/4D^{2}k_{i}^{2}}}\frac{(1-\delta_{i0})\delta_{j0}}{-\mathring{\imath}\omega + r} + \frac{2}{\sqrt{\left(1+v^{2}/4D^{2}k_{i}^{2}\right)\left(1+v^{2}/4D^{2}k_{j}^{2}\right)}}\frac{(1-\delta_{i0})(1-\delta_{j0})}{-\mathring{\imath}\omega + Dk_{j}^{2} + \frac{v^{2}}{4D} + r} \Bigg),
\end{split}
\end{equation}
\begin{equation}\label{eq:MatrixA'}
\begin{split}
\mathsf{A}_{ij}' &= -\frac{\ell\alpha}{L}\Bigg( \frac{\mathrm{Pe}}{1-e^{-\mathrm{Pe}}}\frac{\delta_{i0}\delta_{j0}}{-\mathring{\imath}\omega + r} +  \frac{\mathrm{Pe}}{1-e^{-\mathrm{Pe}}}e^{-\mathrm{Pe}/2}\frac{(-1)^{j}}{\sqrt{1+v^{2}/4D^{2}k_{j}^{2}}}\frac{\delta_{i0}(1-\delta_{j0})}{-\mathring{\imath}\omega + Dk_{j}^{2} + \frac{v^{2}}{4D} + r}\\
&~+ \frac{2e^{\mathrm{Pe}}}{\sqrt{1+v^{2}/4D^{2}k_{i}^{2}}}\frac{(1-\delta_{i0})\delta_{j0}}{-\mathring{\imath}\omega + r} + \frac{2e^{\mathrm{Pe}/2}(-1)^{j}}{\sqrt{\left(1+v^{2}/4D^{2}k_{i}^{2}\right)\left(1+v^{2}/4D^{2}k_{j}^{2}\right)}}\frac{(1-\delta_{i0})(1-\delta_{j0})}{-\mathring{\imath}\omega + Dk_{j}^{2} + \frac{v^{2}}{4D} + r} \Bigg),
\end{split}
\end{equation}
\begin{equation}\label{eq:MatrixB'}
\begin{split}
\mathsf{B}_{ij}' = -\mathsf{D}_{ij}'e^{-\mathrm{Pe}} &= \frac{\ell\alpha}{L}\Bigg( \frac{\mathrm{Pe}}{1-e^{-\mathrm{Pe}}}e^{-\mathrm{Pe}}\frac{\delta_{i0}\delta_{j0}}{-\mathring{\imath}\omega + r} +  \frac{\mathrm{Pe}}{1-e^{-\mathrm{Pe}}}\frac{e^{-\mathrm{Pe}}}{\sqrt{1+v^{2}/4D^{2}k_{j}^{2}}}\frac{\delta_{i0}(1-\delta_{j0})}{-\mathring{\imath}\omega + Dk_{j}^{2} + \frac{v^{2}}{4D} + r}\\
&~+ \frac{2e^{-\mathrm{Pe}/2}(-1)^{i}}{\sqrt{1+v^{2}/4D^{2}k_{i}^{2}}}\frac{(1-\delta_{i0})\delta_{j0}}{-\mathring{\imath}\omega + r} + \frac{2e^{-\mathrm{Pe}/2}(-1)^{i}}{\sqrt{\left(1+v^{2}/4D^{2}k_{i}^{2}\right)\left(1+v^{2}/4D^{2}k_{j}^{2}\right)}}\frac{(1-\delta_{i0})(1-\delta_{j0})}{-\mathring{\imath}\omega + Dk_{j}^{2} + \frac{v^{2}}{4D} + r} \Bigg),
\end{split}
\end{equation}
\begin{equation}\label{eq:MatrixC'}
\begin{split}
\mathsf{C}_{ij}' &= \frac{\ell\alpha}{L}\Bigg( \frac{\mathrm{Pe}}{1-e^{-\mathrm{Pe}}}e^{\mathrm{Pe}}\frac{\delta_{i0}\delta_{j0}}{-\mathring{\imath}\omega + r} +  \frac{\mathrm{Pe}}{1-e^{-\mathrm{Pe}}}\frac{e^{\mathrm{Pe}/2}(-1)^{j}}{\sqrt{1+v^{2}/4D^{2}k_{j}^{2}}}\frac{\delta_{i0}(1-\delta_{j0})}{-\mathring{\imath}\omega + Dk_{j}^{2} + \frac{v^{2}}{4D} + r}\\
&~+ \frac{2e^{\mathrm{Pe}}}{\sqrt{1+v^{2}/4D^{2}k_{i}^{2}}}\frac{(1-\delta_{i0})\delta_{j0}}{-\mathring{\imath}\omega + r} + \frac{2e^{\mathrm{Pe}/2}(-1)^{j}}{\sqrt{\left(1+v^{2}/4D^{2}k_{i}^{2}\right)\left(1+v^{2}/4D^{2}k_{j}^{2}\right)}}\frac{(1-\delta_{i0})(1-\delta_{j0})}{-\mathring{\imath}\omega + Dk_{j}^{2} + \frac{v^{2}}{4D} + r} \Bigg).
\end{split}
\end{equation}

\subsection{\label{app:MatrixMultiplicationEigenvalues}Perturbative matrices closed under multiplication}

The perturbative matrices are, up to multiplicative factors $f$ and $g$, closed under the matrix multiplications that can feature in products of $\mathsf{M}$, Eq.~(\ref{eq:GeometricMatrixPerturbativeMatrices}). The resulting multiplication table is
\setlength{\tabcolsep}{5pt} 
\renewcommand{\arraystretch}{1.5} 
\begin{table}[H]
\begin{center}
\begin{tabular}{ c|c c c c c c c c } 
  & $\mathsf{A}$ & $\mathsf{B}$ & $\mathsf{C}$ & $\mathsf{D}$ & $\mathsf{A}'$ & $\mathsf{B}'$ & $\mathsf{C}'$ & $\mathsf{D}'$ \\
 \hline
 $\mathsf{A}$ & $f\mathsf{A}$ & $g\mathsf{B}'$ &  &  & $g\mathsf{A}$ & $f\mathsf{B}'$ &  &  \\ 
 $\mathsf{B}$ &  &  & $g\mathsf{A}'$ & $f\mathsf{B}$ &  &  & $f\mathsf{A}'$ & $g\mathsf{B}$ \\ 
 $\mathsf{C}$ & $f\mathsf{C}$ & $g\mathsf{D}'$ &  &  & $g\mathsf{C}$ & $f\mathsf{D}'$ &  &  \\ 
 $\mathsf{D}$ &  &  & $g\mathsf{C}'$ & $f\mathsf{D}$ &  &  & $f\mathsf{C}'$ & $g\mathsf{D}$ \\ 
 $\mathsf{A}'$ & $f\mathsf{A}'$ & $g\mathsf{B}$ &  &  & $g\mathsf{A}'$ & $f\mathsf{B}$ &  &  \\ 
 $\mathsf{B}'$ &  &  & $g\mathsf{A}$ & $f\mathsf{B}'$ &  &  & $f\mathsf{A}$ & $g\mathsf{B}'$ \\ 
 $\mathsf{C}'$ & $f\mathsf{C}'$ & $g\mathsf{D}$ &  &  & $g\mathsf{C}'$ & $f\mathsf{D}$ &  &  \\ 
 $\mathsf{D}'$ &  &  & $g\mathsf{C}$ & $f\mathsf{D}'$ &  &  & $f\mathsf{C}$ & $g\mathsf{D}'$ \\ 
\end{tabular}
\end{center}\caption{Multiplication table of the perturbative matrices defined in Appendix \ref{app:PerturbativeMatrixAlgebra}. Rows indicate the matrix multiplying from the left, e.g. $\text{row}=\mathsf{A}$, $\text{column}=\mathsf{B}$ corresponds to $\mathsf{A}\mathsf{B} = g\mathsf{B}'$. Blank entries indicate multiplications that never occur in theory, since the ordering of their corresponding Feynman diagrams prohibits it, see Eq.~(\ref{eq:GeometricMatrixPerturbativeMatrices}).}
\label{table:MatrixMultiplication}
\end{table}

The multiplicative factors resulting from the multiplications have the explicit form
\begin{equation}\label{eq:MatrixEigenvalue_f}
\begin{split}
f(\omega) &= -\frac{\ell\alpha}{L}\left(\frac{vL/D}{1-\exp(-vL/D)} \frac{1}{-\mathring{\imath}\omega + r} + 2 \sum_{j=1}^{\infty}\frac{Dk_{j}^{2}}{(Dk_{j}^{2}+\frac{v^{2}}{4D})(-\mathring{\imath}\omega + Dk_{j}^{2}+\frac{v^{2}}{4D}+r)} \right)\\
&= -\frac{\ell\alpha}{-\mathring{\imath}\omega + r}\left[\sqrt{\frac{-\mathring{\imath}\omega + \frac{v^{2}}{4D}+r}{D}}\coth\left(L\sqrt{\frac{-\mathring{\imath}\omega + \frac{v^{2}}{4D}+r}{D}}\right) + \frac{v}{2D}\right],
\end{split}
\end{equation}
\begin{equation}\label{eq:MatrixEigenvalue_g}
\begin{split}
g(\omega) &= -\frac{\ell\alpha}{L}\left(\frac{vL/D}{1-\exp(-vL/D)} \frac{1}{-\mathring{\imath}\omega + r} + 2e^{\frac{v L}{2D}} \sum_{j=1}^{\infty}\frac{Dk_{j}^{2}(-1)^{j}}{(Dk_{j}^{2}+\frac{v^{2}}{4D})(-\mathring{\imath}\omega + Dk_{j}^{2}+\frac{v^{2}}{4D}+r)} \right)\\
&= -\frac{\ell\alpha}{-\mathring{\imath}\omega + r}\sqrt{\frac{-\mathring{\imath}\omega + \frac{v^{2}}{4D}+r}{D}}\csch\left(L\sqrt{\frac{-\mathring{\imath}\omega + \frac{v^{2}}{4D}+r}{D}}\right)e^{\frac{v L}{2D}},
\end{split}
\end{equation}
where the closed-form representations of the infinite sums can be obtained using techniques for summing over Matsubara frequencies \cite{uzunov1993introduction}.

The ordered structure of the multiplication table \ref{table:MatrixMultiplication} suggests the present theory has applications to more general boundary tumbling phenomena. For instance, we expect to see the same structure for a version of Eq.~(\ref{eq:FokkerPlanckCoupledTumbling}) that has asymmetric self-propulsion speeds, $|v_{L}| \neq |v_{R}|$, and boundary tumbling rates, $\alpha_{L} \neq \alpha_{R}$, for each species. However, in this case, the multiplicative factors would split into non-degenerate values for each of the species, i.e.\ $f \rightarrow f_{L}, f_{R}$ and $g \rightarrow g_{L}, g_{R}$.
\end{widetext}

\subsection{\label{app:PropagatorDerivation}Derivation of propagators}

Using the multiplication table \ref{table:MatrixMultiplication}, the full propagator in Eq.~(\ref{eq:FullPropagatorNotSolved}) can be reduced to closed form by expanding the inverse matrices it contains as geometric series. Firstly, according to the table, $\mathsf{D}\mathsf{D}=f\mathsf{D}$ and therefore $\left(\mathbb{1} - \mathsf{D} \right)^{-1} = \mathbb{1} + (1-f)^{-1}\mathsf{D}$ can be used to rewrite Eq.~(\ref{eq:FullPropagatorNotSolved}) as
\begin{equation}\label{eq:FullPropagatorDerivation1}
\begin{tikzpicture}[dot/.style={fill,circle,inner sep=0pt,outer sep=0pt,minimum size=8pt,label={[label distance=0cm]#1}}, baseline={(current bounding box.center)}]
    \begin{feynman}
    \node [dot] (m1) ;
    \vertex [   left=2 em of m1] (w) ;
    \vertex [      right=2 em of m1] (w0) ;
    \diagram* {
        (w) -- [red, line width=0.5mm, -] (m1) -- [red, line width=0.5mm, -] (w0),
    };
    \vertex [below=1.2 em of m1] {\(\)};
    \end{feynman}
    \end{tikzpicture} = \begin{tikzpicture}[dot/.style={fill,circle,inner sep=0pt,outer sep=0pt,minimum size=10pt,label={[label distance=0.3cm]#1}}, baseline={(current bounding box.center)}]
    \begin{feynman}
    \vertex (w) ;
    \vertex [      right=3em of w] (w0) ;
    \diagram* {
        (w) -- [red, line width=0.5mm, -] (w0),
    };
    \vertex [below=0.3 em of m1] {\(\)};
    \end{feynman}
    \end{tikzpicture}
    \left(\mathbb{1} - \mathsf{B}\mathsf{C} - \frac{1}{1-f}\mathsf{B}\mathsf{D}\mathsf{C} - \mathsf{A}  \right)^{-1}.
\end{equation}
Next, the products $\mathsf{B}\mathsf{C} = g\mathsf{A}'$, $\mathsf{D}\mathsf{C} = g\mathsf{C}'$ and subsequently $\mathsf{B}\mathsf{C}' = f\mathsf{A}'$ give
\begin{equation}\label{eq:FullPropagatorDerivation2}
\begin{tikzpicture}[dot/.style={fill,circle,inner sep=0pt,outer sep=0pt,minimum size=8pt,label={[label distance=0cm]#1}}, baseline={(current bounding box.center)}]
    \begin{feynman}
    \node [dot] (m1) ;
    \vertex [   left=2 em of m1] (w) ;
    \vertex [      right=2 em of m1] (w0) ;
    \diagram* {
        (w) -- [red, line width=0.5mm, -] (m1) -- [red, line width=0.5mm, -] (w0),
    };
    \vertex [below=1.2 em of m1] {\(\)};
    \end{feynman}
    \end{tikzpicture} = \begin{tikzpicture}[dot/.style={fill,circle,inner sep=0pt,outer sep=0pt,minimum size=10pt,label={[label distance=0.3cm]#1}}, baseline={(current bounding box.center)}]
    \begin{feynman}
    \vertex (w) ;
    \vertex [      right=3em of w] (w0) ;
    \diagram* {
        (w) -- [red, line width=0.5mm, -] (w0),
    };
    \vertex [below=0.3 em of m1] {\(\)};
    \end{feynman}
    \end{tikzpicture}
    \left(\mathbb{1} - \left(\mathsf{A} + \frac{g}{1-f}\mathsf{A}' \right)  \right)^{-1}.
\end{equation}
Finally, using Table \ref{table:MatrixMultiplication} for products of $\mathsf{A}$ and $\mathsf{A}'$ results in
\begin{equation}
    \left(\mathsf{A} + \frac{g}{1-f}\mathsf{A}'\right)^{n} = \left(f + \frac{g^{2}}{1-f}\right)^{n-1}\left(\mathsf{A} + \frac{g}{1-f}\mathsf{A}'\right),
\end{equation}
such that expanding the right-hand side of Eq.~(\ref{eq:FullPropagatorDerivation2}) as a geometric series obtains Eq.~(\ref{eq:FullPropagatorFourier}). Eq.~(\ref{eq:FullPropagatorFourierLeftMover}), and the propagators to observe a left-moving particle, can be obtained by a similar procedure.

\section{\label{app:ClosedFormDerivation}Derivation of closed-form expression for $h(x)$}

Here, we show how to go from the series representation of $h(x)$, Eq.~(\ref{eq:InfiniteFourierSeries}), to its closed-form expression in Eq.~(\ref{eq:ClosedFourierSeries}).

First, we know, and can verify numerically before calculating it explicitly in Eq.~(\ref{eq:RightMoverCurrent}), that the current of the right-moving particle $J_{R}$ is constant across space at stationarity. Hence,
\begin{equation}\label{eq:ConstantCurrent}
    \int_{0}^{L}\mathrm{d}x~J_{R}(x) = LJ_{R}.
\end{equation}
The integrand on the left-hand side of Eq.~(\ref{eq:ConstantCurrent}) can be rewritten using the general expression for the current of a right-moving particle, Eq.~(\ref{eq:CurrentNoTumbling}). This relates the current to $P_{R}(x)$ and, in turn, $h(x)$ via Eq.~(\ref{eq:ProbabilityDensityNotNeat}). After inserting the series representation of $h(x)$, Eq.~(\ref{eq:InfiniteFourierSeries}), the left-hand side of Eq.~(\ref{eq:ConstantCurrent}) can be written in closed form after exchanging integration and summation. This gives $vC_{\mathrm{Pe}}/2 = LJ_{R}$, in agreement with Eq.~(\ref{eq:RightMoverCurrent}).

The closed-form expression $J_{R} = vC_{\mathrm{Pe}}/2L$, in combination with Eqs.~(\ref{eq:CurrentNoTumbling}) and (\ref{eq:ProbabilityDensityNotNeat}), can be used to derive the differential equation
\begin{equation}\label{eq:InfiniteSumODE}
    \frac{\mathrm{d}h(x)}{\mathrm{d}x}-\frac{v}{2D}h(x)=-\frac{L}{2D}e^{-\frac{v}{2D}x},
\end{equation}
which is solved by
\begin{equation}\label{eq:InfiniteSumODEResult}
    h(x) = \frac{L}{2v}\left(e^{-\frac{v}{2D}x} + c(\mathrm{Pe}) e^{\frac{v}{2D}x} \right),
\end{equation}
where $c(\mathrm{Pe})$ is a constant yet to be determined. 
To find the explicit form of $c(\mathrm{Pe})$, Eq.~(\ref{eq:InfiniteSumODEResult}) is rearranged to give
\begin{equation}\label{eq:InfiniteSumODEConstant}
    c(\mathrm{Pe}) = \frac{2v}{L}h(x)e^{-\frac{v}{2D}x} -  e^{-\frac{v}{D}x} = \text{const.}
\end{equation}
Then, after inserting the series representation of $h(x)$, Eq.~(\ref{eq:InfiniteFourierSeries}), into Eq.~(\ref{eq:InfiniteSumODEConstant}), one can verify numerically $c(1) \approx -1/(e-1)$ and $c(2) \approx -2/(e^{2}-1)$ etc. Hence, we conjecture $c(\mathrm{Pe}) = -\mathrm{Pe}/(\exp(\mathrm{Pe})-1)$, implying $h(x)$ in Eq.~(\ref{eq:ClosedFourierSeries}) is the closed-form representation of Eq.~(\ref{eq:InfiniteFourierSeries}).

To verify this conjecture, we calculated the coefficients $a_{n} = 2\int_{0}^{L}\mathrm{d}x~h(x)\cos(k_{n}x + \theta_{n})/L$ in the basis expansion $h(x) = \sum_{n=1}^{\infty}a_{n}\cos(k_{n}x + \theta_{n})$, using the conjectured form of $h(x)$, Eq.~(\ref{eq:ClosedFourierSeries}). The calculated coefficients $a_{n}$ indeed match those in Eq.~(\ref{eq:InfiniteFourierSeries}), which verifies that $h(x)$ in Eqs.~(\ref{eq:InfiniteFourierSeries}) and (\ref{eq:ClosedFourierSeries}) are the same function.

\bibliography{references}

\end{document}